\documentclass[fleqn,usenatbib]{mnras}
\usepackage{newtxtext,newtxmath}
\usepackage[T1]{fontenc}
\usepackage{ae,aecompl}

\usepackage{graphicx}	
\usepackage{amsmath}	
\usepackage{amssymb}	

\newcommand{\tmeight}{2M0805$+$48} 
\newcommand{\tmten}{2M1059$-$21} 
\newcommand{\wiseven}{WI0720$-$08} 
\newcommand{\teta}{\boldsymbol{\theta}}

\title[Spectral binary orbits]{Astrometric orbits of spectral binary brown dwarfs I:\\ Massive T dwarf companions to \tmten\ and \tmeight }

\author[J.\ Sahlmann et al.]{
J.\ Sahlmann,$^{1,2}$\thanks{E-mail: josahlmann@gmail.com (JS)}
A.\ J.\ Burgasser,$^{3}$
D.\ C.\ Bardalez Gagliuffi,$^{4}$
P.\ F.\ Lazorenko,$^{5}$\newauthor
D.\ S\'egransan,$^{6}$
M.\ R.\ Zapatero Osorio,$^{7}$
C.\ H.\ Blake,$^{8}$
C.\ R.\ Gelino,$^{9}$
E.\ L.\ Mart\'{i}n,$^{10,11,12}$\newauthor
and H.\ Bouy$^{13}$\\
$^{1}$Independent researcher\\
$^{2}$Space Telescope Science Institute, 3700 San Martin Drive, Baltimore, MD 21218, USA\\
$^{3}$Department of Physics, University of California, San Diego, CA 92093, USA\\
$^{4}$Department of Astrophysics, American Museum of Natural History, Central Park West at 79th St., New York, NY, 10024, USA\\
$^{5}$Main Astronomical Observatory, National Academy of Sciences of the Ukraine, Zabolotnogo 27, 03680 Kyiv, Ukraine\\
$^{6}$Observatoire de l'Universit\'e de Gen\`eve, Chemin de P\'egase 51, 1290 Versoix, Switzerland\\
$^{7}$Centro de Astrobiolog\'{i}a (CSIC-INTA), Carretera Ajalvir km 4, E-28850 Torrej\'{o}n de Ardoz, Madrid, Spain\\
$^{8}$University of Pennsylvania, Department of Physics and Astronomy, Philadelphia, PA 19104, USA\\
$^{9}$NASA Exoplanet Science Institute, Mail Code 100-22, California Institute of Technology, 770 South Wilson Avenue, Pasadena, CA 91125, USA\\
$^{10}$Instituto de Astrof\'isica de Canarias (IAC), Calle V\'ia L\'actea s/n, E-38200 La Laguna, Tenerife, Spain\\
$^{11}$Departamento de Astrof\'isica, Universidad de La Laguna (ULL), E-38206 La Laguna, Tenerife, Spain\\
$^{12}$Consejo Superior de Investigaciones Cient\'{i}­ficas, E-28006 Madrid, Spain\\
$^{13}$Laboratoire d'astrophysique de Bordeaux, University of Bordeaux, CNRS, B18N, All\'{e}e Geoffroy Saint-Hilaire, 33615 Pessac, France\\
}

\date{Accepted 2020 April 30. Received 2020 April 28; in original form 2020 January 31.}

\pubyear{2020}

\begin{document}
\label{firstpage}
\pagerange{\pageref{firstpage}--\pageref{lastpage}}
\maketitle

\begin{abstract}
Near-infrared spectroscopic surveys have uncovered a population of short-period, blended-light spectral binaries composed of low-mass stars and brown dwarfs. These systems are amenable to orbit determination and individual mass measurements via astrometric monitoring. Here we present first results of a multi-year campaign to obtain high-precision absolute astrometry for spectral binaries using the Gemini-South and Gemini-North GMOS imagers. We measure the complete astrometric orbits for two systems: {\tmeight} and {\tmten}. Our astrometric orbit of {\tmeight} is consistent with its 2-year radial velocity orbit determined previously and we find a mass of $66^{+5}_{-14} M_\mathrm{Jup}$ for its T5.5 companion.
For \tmten\ we find a 1.9 year orbital period and a mass of $67^{+4}_{-5} M_\mathrm{Jup}$ for its T3.5 companion. 
We demonstrate that sub-milliarcsecond absolute astrometry can be obtained with both GMOS imagers and that this is an efficient avenue for confirming and characterising ultracool binary systems.
\end{abstract}


\section{Introduction}

The internal and observable properties of normal hydrogen-burning stars can be determined largely from their masses, ages, and compositions \citep{Vogt:1926aa,Russell:1931aa}. This is not the case for brown dwarfs, low-mass sources incapable of sustained hydrogen fusion \citep{Hayashi:1963uq,Kumar:1963zr}, which were finally identified in the Pleiades cluster just 25 years ago \citep{1995Natur.377..129R}. Because these objects cool over time, both mass and age dictate observable properties, challenging the characterization of the local brown dwarf population and substellar mass function \citep{Burgasser:2004ab,2005ApJ...625..385A}. Disentangling mass and age is a primary motivator for characterizing spectral features sensitive to surface gravity \citep{Allers:2007aa,Cruz:2009fk,2010A&A...517A..53M} which are, however, too subtle for sources older than $\sim$200 Myr \citep{Kirkpatrick:2008uk,2017ApJ...838...73M}. Other stellar age metrics such as magnetic activity and angular momentum evolution are not useable in the brown dwarf regime \citep{Mohanty:2003aa,Berger:2006aa,Reiners:2008cr}. As a result, we cannot precisely measure the ages or masses of the majority of brown dwarfs in the vicinity of the Sun. In many cases, e.g.\ for L dwarfs or when the Lithium test \citep[e.g.][]{Magazzu:1993kx} is inconclusive, we cannot even determine if they are brown dwarfs or stars, although recent progress is being made by extending the determination of the lithium boundary method L dwarf members in clusters older than it was previously thought to be possible, such as the Hyades \citep[e.g.][]{2018ApJ...856...40M}.

Over the past decade, mass measurements have been achieved for dozens of low-mass stellar and sub-stellar binaries in the field \cite[cf.][]{2017ApJS..231...15D}. These are primarily resolved systems, but there have also been a smaller number of mass measurements from radial velocity \citep[RV; ]{Basri:1999kx,2003A&A...401..677G,2007ApJ...666L.113J,2010A&A...521A..24J,Blake:2008fr,2010ApJ...723..684B,Burgasser:2010kx,2012ApJ...757..110B,Burgasser:2016aa,Konopacky:2010dk} and astrometric orbits \citep{Sahlmann:2015ac,Sahlmann:2013ab,2016AJ....151...57K}, as well as microlensing masses \citep[e.g.][]{Bennett:2008ys, Han:2013aa, 2018AJ....156..136M,2017A&A...604A.103P}. The direct measurement of both spectra and masses for resolved short-period binaries have permitted tests of evolutionary models, revealing in many cases systematic discrepancies \citep[e.g.][]{Konopacky:2010dk,Dupuy:2014aa,2018ApJ...865...28D,2019arXiv191001652B}. Yet the number of resolvable systems (angular separation $\gtrsim$50 mas equivalent to $\gtrsim1$AU at 25 pc) with short enough orbital periods for mass measurement ($P \lesssim 10$ yr) has reached a limit until the resolving power of 30m telescopes becomes available. Other methods yield fewer binaries and less information: radial-velocity (RV) monitoring ($\lesssim10$\% yield) probes shorter orbits but provide only mass limits because of the unknown orbit inclination and little information on secondary atmospheres. Astrometric variables are rarer ($\lesssim5$\% yield) and provide limited information on the secondary spectrum. Microlenses offer precise masses but no atmospheric information whatsoever.

Fortunately, close-separation, unequal-mass binaries straddling the hydrogen-burning limit can be identified as blended-light spectral binaries \citep{2004ApJ...604L..61C,2008ApJ...681..579B,Burgasser:2010kx, Bardalez-Gagliuffi:2014aa_}. These systems are composed of a late-M or L dwarf primary and a T dwarf secondary, which have distinct spectral morphologies \citep{2005ARA&A..43..195K}. They are efficiently revealed through peculiar features in low-resolution, near-infrared spectra, which can also be used to characterize the atmospheres of the binary components. Their separation-independent identification permits the detection and orbital measurement of potentially short-period systems. Over 60 spectral binary candidates have been identified to date, and roughly a dozen confirmed through high resolution imaging, RV monitoring, and astrometric monitoring (\citealt{Bardalez-Gagliuffi:2015aa} and references therein). These include some of the most tightly-separated very low-mass binaries known, for which both orbit and mass measurements have been achieved, primarily through RV monitoring \citep[$<$1 AU, ][]{Blake:2008fr,Burgasser:2008uq,Burgasser:2012aa,Burgasser:2016aa}

However, RV monitoring typically provides only one axis of the primary's orbital motion. Fortunately, the proximity and extreme flux ratios of very low-mass spectral binaries make them ideal for the measurement of astrometric variability and therefore all orbital parameters \citep[e.g.][]{Brandner:2004aa, Dahn:2008fk, Dupuy:2012fk, Sahlmann:2013ab}.
Here we report first results from a long-term, ground-based astrometric follow-up survey targeting very low-mass spectral binaries.

\section{Astrometric Binary Survey sample}\label{sec:targets}
The amplitude of astrometric variability, i.e.\ the size of a binary's photocentre orbit $\alpha = a_\mathrm{rel} \, (f-\beta)$, 
depends on the projected angular semimajor axis of the orbit $a_\mathrm{rel}$, the fractional mass $f = M_2 / (M_1+M_2)$ and the fractional flux $\beta = F_2 / (F_1+F_2) = (1+10^{0.4 \Delta m})^{-1}$, where $\Delta m$ is the magnitude difference between the components in a given photometric band.
Very low-mass spectral binaries, particularly those with late-M and early-L primaries, can have very large optical flux ratios ($F_2 / F_1  \ll 1 \Rightarrow \beta \ll 1$) and modest mass ratios ($q = M_2/M_1 \simeq 0.5-0.8 \Rightarrow f \simeq 0.5-0.7$), depending on the system age \citep{Burgasser:2009fk}. Hence, spectral binaries should in principle exhibit non-zero astrometric perturbations. Detectable astrometric variables must strike a balance between large-amplitude but long-period systems that could be resolved by direct imaging ($\alpha \gtrsim 20-50$ mas for $a_\mathrm{rel} > 50-100$ mas; $P \gtrsim 3-8$ yr); and short-period but small-amplitude systems that could be identified by RV monitoring ($\alpha \lesssim 3-5$ mas for $a_\mathrm{rel} \lesssim 0.3$ AU at 25 pc; $P \lesssim0.5$ yr; RV signature $\lesssim 5-10$ km/s). Hence, our optimal targets should be unresolved by direct imaging, and may or may not show some evidence of RV variability, encompassing periods of 0.5 yr $\lesssim P \lesssim 3$ yr.

Starting from the 60 spectral binaries compiled by \citet{Bardalez-Gagliuffi:2014aa_} and subsequent discoveries, we selected sources within 40 pc for which high-resolution images were unresolved, and for which the inferred $\Delta J > 1$ (based on modeling of the blended spectrum), implying $\Delta I \gtrsim 4$ based on the $I-J$/spectral type relation of \citet{Hawley:2002aa}. We also excluded sources with $I > 22$. 
We prioritized sources for which RV variations are evident in follow-up Keck/NIRSPEC monitoring (\citealt{Burgasser:2016aa}; Burgasser et al.\ in prep.). 
The final sample consists of 10 spectral binaries and binary candidates, listed in Table \ref{tab:targets}.

These targets were observed in three long-term astrometric monitoring campaigns. Eight sources (four in the North and four in the South) were monitored with the Gemini Multi-Object Spectrograph (GMOS; \citealt{2004PASP..116..425H,2016SPIE.9908E..2SG}) on the Gemini-North\footnote{Programs GN-2017A-Q-24, GN-2017B-Q-4, GN-2018A-Q-128, GN-2018B-Q-104, GN-2019A-Q-231, GN-2019B-Q-104} and -South\footnote{Programs GS-2015A-Q-69, GS-2015B-Q-2, GS-2016B-Q-36, GS-2017A-Q-53, GS-2017B-Q-6, GS-2018A-Q-133} telescopes (PI Burgasser). Two sources (SDSS~J0931$+$28 and 2MASS~J1453$+$14) were observed with OSIRIS \citep{2000SPIE.4008..623C} on the Gran Telescopio Canaria (GTC; PI Sahlmann). In this paper, we focus on the results of our GMOS observations of {\tmeight} and {\tmten}; the remaining targets will be discussed in subsequent publications.

\begin{table*}
\centering
\caption{Target sample.}
\label{tab:targets}
\begin{tabular}{lllccccll}
\hline
Name & Identifier & Combined &  Component & $I$ & $\Delta{J}$ & $\Delta{I}$ & Instrument & Ref \\
 & & Type & Types & (mag) & (mag) & (mag) \\
\hline
\tmeight & \href{http://simbad.u-strasbg.fr/simbad/sim-basic?Ident=SDSS J080531.84\%2B481233.0&submit=SIMBAD+search}{SDSS J080531.84+481233.0} & L4/L9 & L4 + T5.5 & 19.59$\pm$0.03 & 1.5 & 3.8 &  Gemini-N/GMOS & 1--4 \\
2M1311$+$36 & \href{http://simbad.u-strasbg.fr/simbad/sim-basic?Ident=2MASS+J13114227\%2B3629235&submit=SIMBAD+search}{2MASS J13114227+3629235} & L5p & L5 + T4 & 20.50$\pm$0.04 & 2.2 & 4.3 & Gemini-N/GMOS & 5 \\
2M1711$+$22 & \href{http://simbad.u-strasbg.fr/simbad/sim-basic?Ident=2MASSI J1711457\%2B223204&submit=SIMBAD+search}{2MASSI J1711457+223204} & L6.5/L9.5 & L5 + T5.5 & 21.92$\pm$0.15 & 0.9 & 3.3 &  Gemini-N/GMOS & 2,5 \\
2M2126$+$76 & \href{http://simbad.u-strasbg.fr/simbad/sim-basic?Ident=2MASS J21265916\%2B7617440&submit=SIMBAD+search}{2MASS J21265916+7617440} & T0p & L8.5 + T4.5 & 20.13$\pm$0.03 & 0.4 & 1.9 &  Gemini-N/GMOS & 6 \\
\hline
WI0720$-$08 & \href{http://simbad.u-strasbg.fr/simbad/sim-basic?Ident=WISE J072003.20-084651.2&submit=SIMBAD+search}{WISE J072003.20-084651.2} & M9.5 & M9.5 + T5 & 14.93$\pm$0.02 & 2.6 & 5.5 & Gemini-S/GMOS & 7--9 \\
\tmten & \href{http://simbad.u-strasbg.fr/simbad/sim-basic?Ident=2MASSI J1059513-211308&submit=SIMBAD+search}{2MASSI J1059513-211308} & L1/L2  & L0.5 + T3.5 & 21.54$\pm$0.08 & 2.6 & 5.1 &  Gemini-S/GMOS & 5 \\
WI1623$-$05 & \href{http://simbad.u-strasbg.fr/simbad/sim-basic?Ident=WISE J16235970-0508114&submit=SIMBAD+search}{WISE J16235970-0508114} & L1 & L0.5 + T6 & 19.32$\pm$0.01 & 3.4 & 6.4 &  Gemini-S/GMOS & 5\\
2M2026$-$29 & \href{http://simbad.u-strasbg.fr/simbad/sim-basic?Ident=2MASS J20261584-2943124&submit=SIMBAD+search}{2MASS J20261584-2943124} & L1  & L0 + T6 & 19.13$\pm$0.02 & 3.4 & 6.5 &  Gemini-S/GMOS & 5,10 \\
\hline
SD0931$+$28 & \href{http://simbad.u-strasbg.fr/simbad/sim-basic?Ident=SDSS+J093113.23\%2B280227.1&submit=SIMBAD+search}{SDSS J093113.23+280227.1} & L3  &  L1.5 + T2.5 & 19.44$\pm$0.03 & 2.2 & 4.4 &  GTC/OSIRIS & 5 \\
2M1453$+$14 & \href{http://simbad.u-strasbg.fr/simbad/sim-basic?Ident=2MASS+J14532582\%2B1420410&submit=SIMBAD+search}{2MASS J14532582+1420410} & L1 & L1 + T6 & 19.65$\pm$0.02 & 3.3 & 6.2 & GTC/OSIRIS & 5 \\
\hline
\end{tabular}
{\newline References: (1) \citet{Burgasser:2007aa}; (2) \citet{Burgasser:2010kx}; (3) \citet{Dupuy:2012fk}; (4) \citet{Burgasser:2016aa}; (5) \citet{Bardalez-Gagliuffi:2014aa_}; (6) \citet{Bardalez-Gagliuffi:2015aa}; (7) \citet{Burgasser:2015aa}; (8) \citet{2015AJ....150..180B}; (9) \citet{Dupuy:2019aa}; (10) \citet{2010AJ....140..110G} }
\end{table*}

\begin{description}
\item[{\tmten}:] This source has an L1 near-infrared, combined-light classification \citep{Cruz:2003aa}, yet it was identified as a candidate spectral binary of L1+T3 dwarf components \citep{Bardalez-Gagliuffi:2014aa_}.

\item[{\tmeight}:] This source was first identified as an unusually blue L dwarf with discrepant optical and near-infrared classifications of L4 and L9 (\citealt{Hawley:2002aa} and \citealt{Knapp:2004aa}, respectively). \cite{Burgasser:2007aa} posited that its unusual near-infrared spectrum 
could be due to the combined light of a L4.5 and T5 dwarf components. \cite{Dupuy:2012fk} confirmed this source as an astrometric variable
with an amplitude of $\sim$ $15$\,mas, and estimated a period of $2.7-9.1\,$yr and a semi-major axis of $0.9-2.3\,$AU.
They also inferred a similar spectral component composition of L4 + T5.
 High-resolution laser guide-star adaptive-optics imaging observations with Keck were unable to resolve this system \citep{Bardalez-Gagliuffi:2015aa}, setting a separation upper limit of 170\,mas. 
 \citet{Burgasser:2016aa} monitored this system with high resolution infrared spectroscopy, and detected significant RV variability over the course of 4 years.
 Their orbit fits yielded a period of  2.02$\pm$0.03 years, a semimajor axis of 0.76$_{-0.06}^{+0.05}$~au, and a nonzero eccenticity of 0.46$\pm$0.05.
 By combining their measurements with brown dwarf evolutionary models, \citet{Burgasser:2016aa}
 also deduced that the system was close to edge-on (90$\degr\pm$19$\degr$) and has a large system mass ratio ($q$ = ${0.86}_{-0.12}^{+0.10}$), substellar-mass components (M$_1$ = ${0.057}_{-0.014}^{+0.016}$~M$_{\odot}$, M$_2$ = ${0.048}_{-0.010}^{+0.008}$~M$_{\odot}$), and a relatively old age ($\tau$ $\gtrsim$  4~Gyr), although these values are highly model-dependent.
 \end{description}

\section{Observations}
\subsection{Gemini GMOS imaging astrometry}
Our program is modeled on the astrometric survey of ultracool dwarfs described in \cite{Sahlmann:2013ab, Sahlmann:2014aa}, which demonstrated 0.1 mas accuracy for M8--L2.5 dwarfs down to $I = 17.5$ using the VLT/FORS2 instrument (126 mas/pixel) and the method of \cite{Lazorenko:2014aa}. We conducted a pilot program in 2015 to test the application of this method with Gemini-S/GMOS, whose Hamamatsu detector has comparable field-of-view, more red sensitivity, and a smaller pixel scale (80 mas) than VLT/FORS2.
Upon verifying that comparable astrometric precision could be achieved, we conducted monitoring programs with Gemini-S from 2016-2018 and with Gemini-N (after the installation of the Hamamatsu  detector for GMOS-N) from 2017-2019. Table \ref{tab:frame} summarises the data analysed here.

The observation design was straightforward. Multiple epochs of imaging separated by roughly one month during visibility periods were obtained over several years.  Each epoch consisted of a sequence of 8--12 dithered (1\arcsec\ random dither pattern) exposures in the $i$-band filter, with exposure times designed to yield $S/N > 50$ per exposure. The field of view centering and orientation were maintained to be as constant as possible throughout the monitoring period, and observations were obtained at small airmass ($<$1.5) and close to meridian to reduce differential colour refraction (DCR). Reasonable imaging conditions ($<$0.75\arcsec) and sky transparency (cirrus only or better) were also required. All observations were requested and executed in queue mode.

\begin{table}
\centering
\caption{Gemini data analysed in this paper. The individual frame exposure time (in seconds), typical number of dithers per epoch $N_d$, number of epochs $N_e$, total number of frames $N_f$, and number of available reference stars $N_\star$ are listed, as well as the total timespan covered by the observations (in days).}
	\label{tab:frame}
\begin{tabular}{rrrrrrr}
\hline
Source & Exp. & $N_d$ & $N_e$ &$N_f$& Timespan & $N_\star$ \\
 & (s) & & & & (day) & \\
\hline
\tmeight & 200 & 10 & 13 & 125 & 786 & 67 \\
\tmten & 180 & 12 & 19 & 258 & 1153 & 51 \\
\hline
\end{tabular}
\end{table}

\subsection{Keck NIRSPEC radial velocities}
We obtained new high-resolution spectroscopy of {\tmeight} and {\tmten} with Keck/NIRSPEC \citep{1998SPIE.3354..566M} at multiple epochs. The data acquisition, reduction, and RV determination were performed as described in e.g.\ \citet{Burgasser:2012aa,Burgasser:2016aa}.
The resulting measurements are listed in Table~\ref{tab:rv}, and add to the existing measurements for {\tmeight} reported in \citet{Burgasser:2016aa}.

\begin{table}
    \caption{Additional Keck NIRSPEC radial velocity measurements of {\tmeight} and \tmten.}
    \label{tab:rv}
    \centering
\begin{tabular}{lllrr}
    \hline
    \hline
Date (UT) & MJD & S/N & RV  & $v\sin{i}$    \\
&  &  & (km/s) & (km/s) \\
    \hline
{\tmeight} \\
    \hline
2017 Mar 22 &  57834.35930 & 19 &  +7.7$\pm$0.5 & 36.4$\pm$0.8 \\
2017 Dec 7 & 58094.63541 & 11 & +14.1$\pm$0.8 & 35.4$\pm$1.2 \\
2018 Jan 1 & 58119.49976 & 4 & +17.3$\pm$1.1 & 33.9$\pm$2.4 \\
    \hline
{\tmten} \\
    \hline
2016 Apr 22 & 57500.30332 & 18 & +40.0$\pm$0.3 & 13$\pm$2.2 \\
2016 May 22 & 57530.26493 & 16 & +40.0$\pm$0.3 & 14.6$\pm$1.6\\
    \hline
\end{tabular}
\end{table}

\section{Astrometric data reduction and analysis}    
\subsection{Basic data reduction}
We used the Gemini data reduction package\footnote{https://www.gemini.edu/sciops/data-and-results/processing-software} to perform the bias and flatfield corrections of the GMOS images. The identification of applicable calibration data and the generation of master bias and master flat files were automated using python scripts that interface with the Gemini Archive \citep{johannes_sahlmann_2019_3515528}\footnote{\url{https://github.com/Johannes-Sahlmann/gemini-reduction}}. 

\subsection{Source extraction and astrometric analysis}
We used Source Extractor \citep{Bertin:1996aa} and PSF Extractor \citep{Bertin:2006aa} to identify sources in every image frame and determine their pixel positions. After a preliminary source extraction, the PSF Extractor tool was run to generate an empirical PSF with parameters that vary across the field. Then, Source Extractor was provided with the PSF model and performs the final extraction including the pixel position determination. The inter-frame identification of sources was performed using SCAMP \citep{Bertin:2006aa}, however, the astrometric information provided by that tool was not used.

\subsection{Absolute astrometry in the Gaia reference frame}
We based the absolute astrometric calibration of our images on the second Gaia data release \citep[GDR2,][]{Gaia-Collaboration:2018ae,GaiaCollaboration:2016aa}. We accessed the Gaia archive using \texttt{pygacs}\footnote{\url{https://github.com/Johannes-Sahlmann/pygacs}} and downloaded sources from the \emph{gaiadr2.gaia\_source} table within a 4\arcmin\ radius of the target. To increase the fidelity of Gaia sources, we selected only entries with \emph{astrometric\_excess\_noise} $<2$\,mas, \emph{duplicated\_source}$=0$, \emph{ra\_error}$<3$\,mas and \emph{dec\_error}$<3$\,mas. We then computed the source positions and their uncertainties at the reference epoch, whose determination is described below. We took into account the five parameters of the standard astrometric model (positions, parallax, and proper motions) and their covariances when applicable.

To identify the reference frame for absolute alignment, we crossmatched the GDR2 sources with the sources extracted in every frame. The crossmatch includes an iterative approach in which a two-dimensional distortion model (a bivariate polynomial of degree 4) is fit at every iteration and used to continuously refine the position of extracted sources, implemented in \texttt{pystortion} \citep{johannes_sahlmann_2019_3516268}\footnote{Using \url{https://github.com/spacetelescope/pystortion}}. We chose the reference epoch as the image sequence with a large number of crossmatched sources and stable distortion parameters. Within the reference epoch, we chose the reference frame as the one having the largest number of crossmatched sources.

Finally, we performed the absolute astrometric calibration of the reference frame by fitting a two-dimensional distortion model between the GDR2 source positions that were tangent-plane projected with a reference point located at the center of the field, and the pixel positions of extracted sources. This reference frame  defines the absolute coordinates of the measured sources in two ways: it determines the camera distortion and it ties the pixel-based coordinates to absolute stellar coordinates.
Table \ref{tab:referenceframes} lists the reference frame characteristics for the two targets. 
\begin{table*}
\centering
\caption{Reference frames used for absolute astrometric alignment. The reference frame corresponds to the archive file name which encodes the date and the sequence number, $k$ indicates the degree $d=k/2-1$ of the polynomial fit \citep{Lazorenko:2004cs}, $r_\mathrm{x}$ is the initial crossmatch radius, $N_\mathrm{Gaia}$ is the number of high-fidelity GDR2 sources within 4$\arcmin$ of the target, $N_\mathrm{image}$ is the number of extracted sources in the reference frame, $N_\mathrm{x}$ is the number of crossmatched sources used for the absolute alignment, and the last four columns list the pixel scales at the reference point and their fit residuals in both dimensions.}
\label{tab:referenceframes}
\begin{tabular}{rrrrrrrrr@{$\pm$}lr@{$\pm$}lrrrr}
\hline
object & Instrument & Reference frame & $k$ & $r_\mathrm{x}$  & $N_\mathrm{Gaia}$ & $N_\mathrm{image}$ &$N_\mathrm{x}$ & \multicolumn{2}{c}{Scale $x$} & \multicolumn{2}{c}{Scale $y$} & rms $x$ & rms $y$ \\
 & & & & (\arcsec) & & & & \multicolumn{2}{c}{(mas)} & \multicolumn{2}{c}{(mas)} & (mas) & (mas)\\
\hline
\tmten & GMOS-S & S20150216S0113 & 10 & 4.0 & 74 & 114 & 36 & 80.092 & 0.002 & 80.090 & 0.003 & 3.1 & 2.6 \\
\tmeight & GMOS-N & N20171102S0357 & 10 & 4.0 & 49 & 177 & 22 & 80.909 & 0.033 & 80.937 & 0.020 & 1.3 & 1.2 \\
\hline
\end{tabular}
\end{table*}

\subsection{Correction of atmospheric image motion and variable distortion}\label{sec:reduction}
After application of the absolute astrometric calibration of the reference frame onto the International Celestial Reference Frame (ICRF) realized by Gaia DR2, we determined the 2D polynomial transformation between every frame and the reference frame. In contrast to the absolute calibration, we can use all extracted stars in the image frames (in practice, some sources are discarded as part of the process as described below). We applied the methods of \cite{Lazorenko:2014aa,Lazorenko:2009ph} in a slightly simplified fashion.

First, we applied the polynomial transform that was determined for the absolute alignment of the reference frame to all frames of each target field. This step transforms pixel positions into tangent-plane projected angular coordinates that are still affected by changes in the telescope pointing/orientation, variable distortion, and atmospheric image motion.

\begin{figure*}
\includegraphics[width=\textwidth]{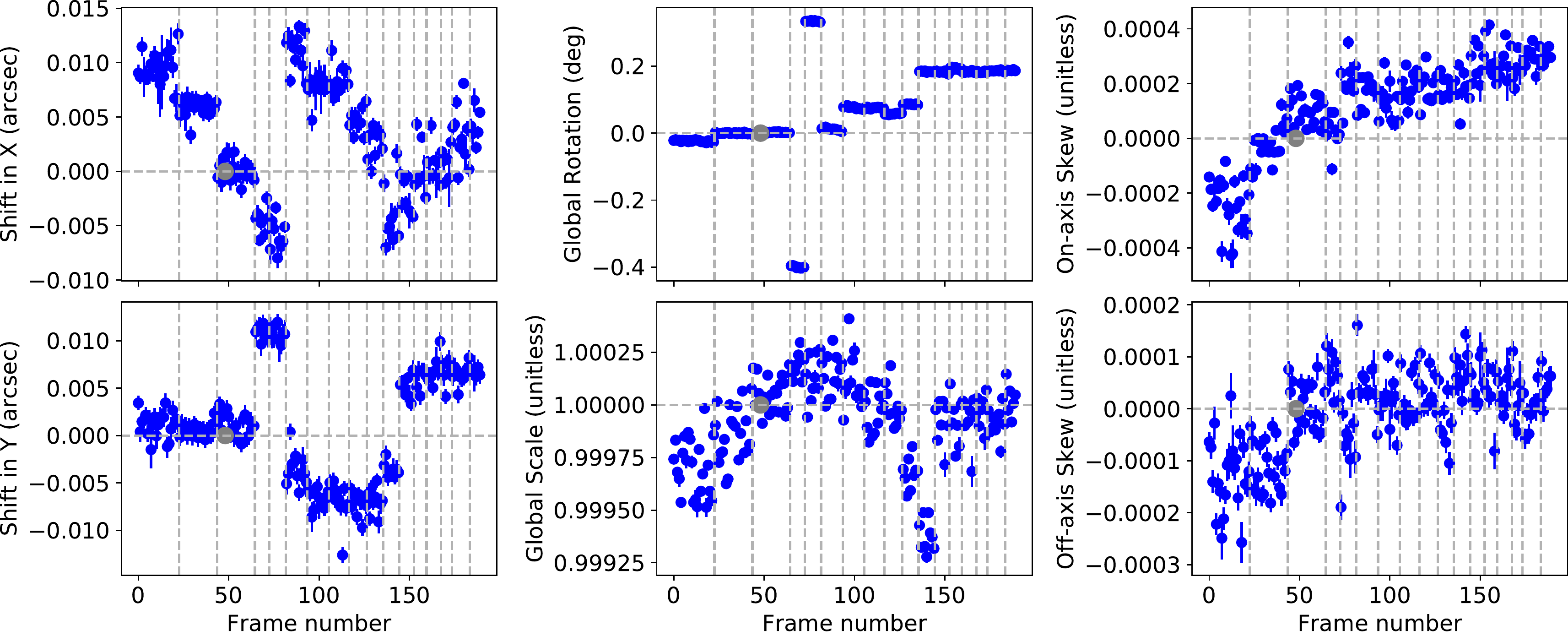}
\caption{Evolution of inter-frame distortion parameters in the \tmten\ field as the result of the iterative reduction described in Section \ref{sec:reduction}. Blue symbols show the frame-by-frame evolution relative to the reference frame indicated by the larger grey circle, which is tied to the Gaia system. Vertical dashed lines indicate the epoch boundaries. \emph{Left}: Residual lateral offsets (these are not the telescope pointing errors because these parameters are being minimized by the procedure). \emph{Middle}: Global rotation and scale. The rotation figure shows the variation in position angle relative to the reference frame.  \emph{Right}: On-axis skew (i.e.\ the scale difference between the axes) and the off-axis skew (i.e.\ the non-perpendicularity between the axes).}
\label{fig:dist_params}
\end{figure*}

Second, we determined the 2D polynomial transformation between each frame and the reference frame for a given target. These transformations correct for the three effects mentioned above, and yield absolute astrometry in every frame for all measured sources. The target is excluded in this step because it is not used to define the astrometric reference field; instead, the target's motion is determined relative to the astrometric frame defined by the reference stars.
In Figure~\ref{fig:dist_params}, we show the evolution of the six first-order distortion parameters (offsets, scale, rotation, and skew terms) in the {\tmten} field, for which $50$ reference stars passed all the iterative filter stages and a 4th degree bivariate polynomial was used to map every frame to the reference frame. The residual root mean square (RMS) of the fits was typically 5--10 mas for both axes.
Changes in the telescope's reported position angle can be seen in the rotation term, and scale and skew changes with amplitudes\footnote{This value are unitless because all positions are measured in arcseconds relative to the reference position.} up to $7\cdot 10^{-4}$ were measured. The sudden scale change around frame 150 corresponds to the time gap between the 2017 and 2018 seasons. The scale difference between the two axes of the GMOS-S detector exhibits a steady increase, whereas the non-perpendicularity between the axes appears to stabilise over time.

The third step consisted of fitting the per-epoch astrometry of individual sources with the appropriate model. The astrometric measurements of the target are $\alpha^{\star}_m= \alpha \cos{\delta}$ and $\delta_m$, corresponding to Right Ascension and Declination, respectively, in frame $m$ at time $t_m$ relative to the reference frame of background stars. These are modeled with six free parameters $\Delta\alpha^{\star}_0, \Delta\delta_0, \mu_{\alpha^\star}, \mu_\delta, \varpi$, and $\rho$ as:
\begin{equation}\label{eq:axmodel}
\begin{array}{ll@{\hspace{2mm}}l}
\!\alpha^{\star}_{m} =& \Delta \alpha^{\star}_0 + \mu_{\alpha^\star} \, t_m + \varpi \, \Pi_{\alpha,m} &-\;\; \rho\, f_{1,x,m} \\
\delta_{m} = &{\Delta \delta_0 + \,\mu_\delta      \,  \;                      t_m \;+ \varpi \, \Pi_{\delta,m}}  &{+\;\; \rho \,f_{1,y,m}},
\end{array}
\end{equation}
where $\Delta\alpha^{\star}_0, \Delta\delta_0$ are the coordinate offsets, $\mu_{\alpha^\star}, \mu_\delta$ are the proper motions, and the parallactic motion is expressed as the product of relative parallax $\varpi$ and the parallax factors $\Pi_\alpha, \Pi_\delta$. The atmospheric refraction {modelled by $\rho$} in Eq.\ (\ref{eq:axmodel}) has one parameter less than the model used for our FORS2 work \citep{Lazorenko:2011lr, Sahlmann:2014aa} because neither Gemini-South nor Gemini-North incorporates a dispersion compensator. In this case, the differential chromatic refraction (DCR) is modeled with the free parameter $\rho$ and the {coefficient} $f_{1}$, where {the latter is fully determined as} a function of zenith angle, temperature, and pressure \citep{Lazorenko:2006qf, Sahlmann:2013ab, Sahlmann:2016aa}. {The DCR treatment does not involve the estimation of source colours, instead $\rho$ is an empirical free model parameter that corresponds to the effective colour of the target relative to the average reference star.}

This reduction technique is applicable equally both to the target and any field star with its unique set of reference stars. Astrometric solutions for field stars were used to test the presence of systematic errors, to derive the parallaxes of background stars, to corrected to absolute parallax, to determine the pixel scale, and to compile the catalogue of field stars. 

As discussed in \citet[][Sect. 3.3]{Lazorenko:2009ph}, the final solution of this third step must be found by iterating steps two and three while imposing an additional set of constraints on the astrometric parameters of the reference stars \citep[see also Sect. 4.1 of][]{Sahlmann:2012fk2}. Essentially, this procedure allows us to determine the astrometric parameters (e.g.\ relative parallax and proper motion) of reference stars while also correcting for the associated epoch-dependent displacements from a rigid reference field against which the target's motion can be measured. The primary constraint imposed on this iterative process is that the sum of astrometric parameters across reference stars vanishes; i.e., that the sum of their relative parallaxes is zero. 

\begin{figure*}
\includegraphics[width=0.7\linewidth]{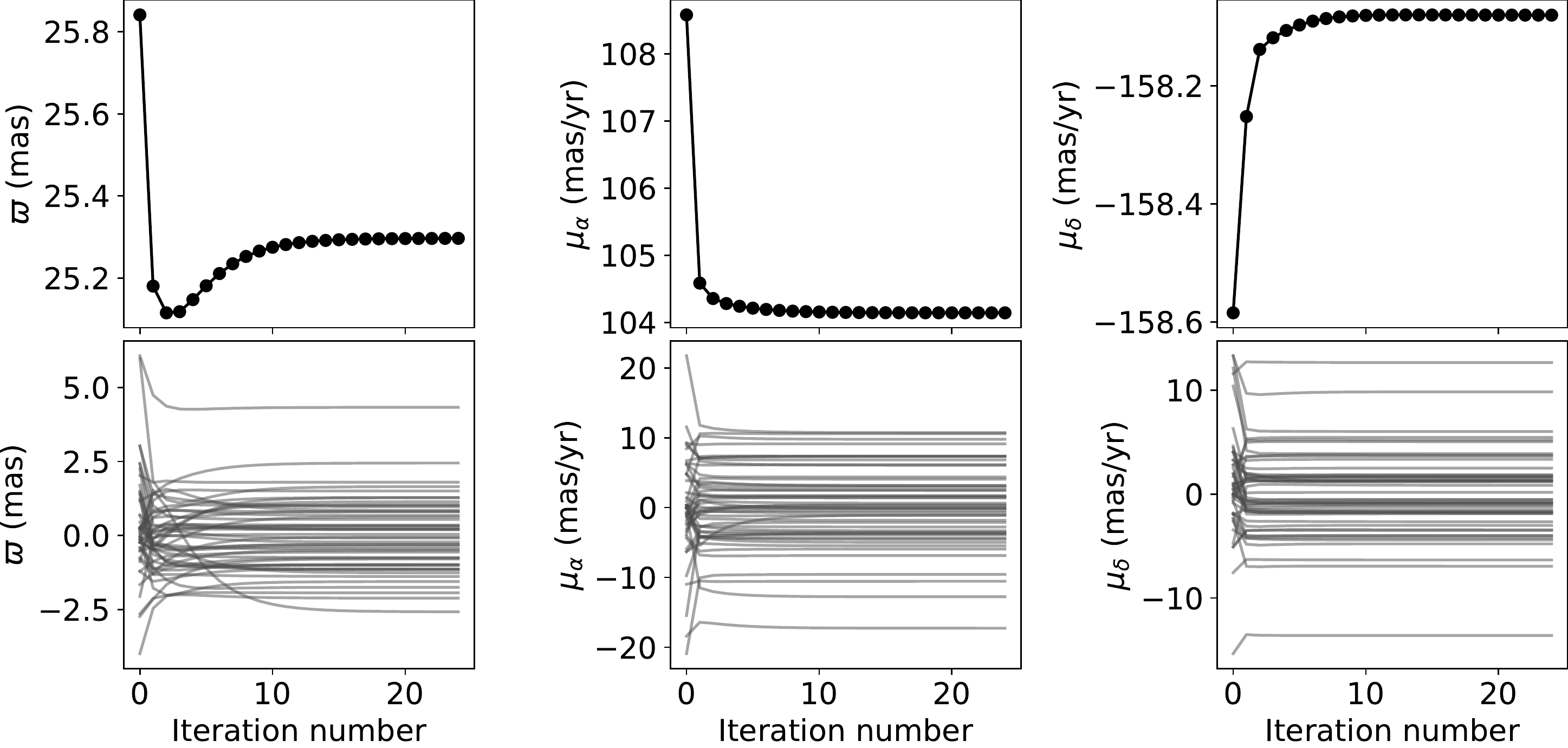}
\caption{Effects of the iterative analysis for \tmten. The evolution of the best-fit parallax (left), proper motion in RA (middle) and Dec (right) is shown as a function of iteration number for the target (top row) and the reference stars (bottom row).}
\label{fig:iterations}
\end{figure*}

Figure \ref{fig:iterations} shows the evolution of the parallax and proper motion parameters for \tmten as a function of iteration number. It demonstrates the significant benefit of using this approach and the need for 15--20 iterations to allow the algorithm to converge.

\subsection{Correction to absolute parallax and proper motion}
\begin{figure}
\includegraphics[width=\columnwidth]{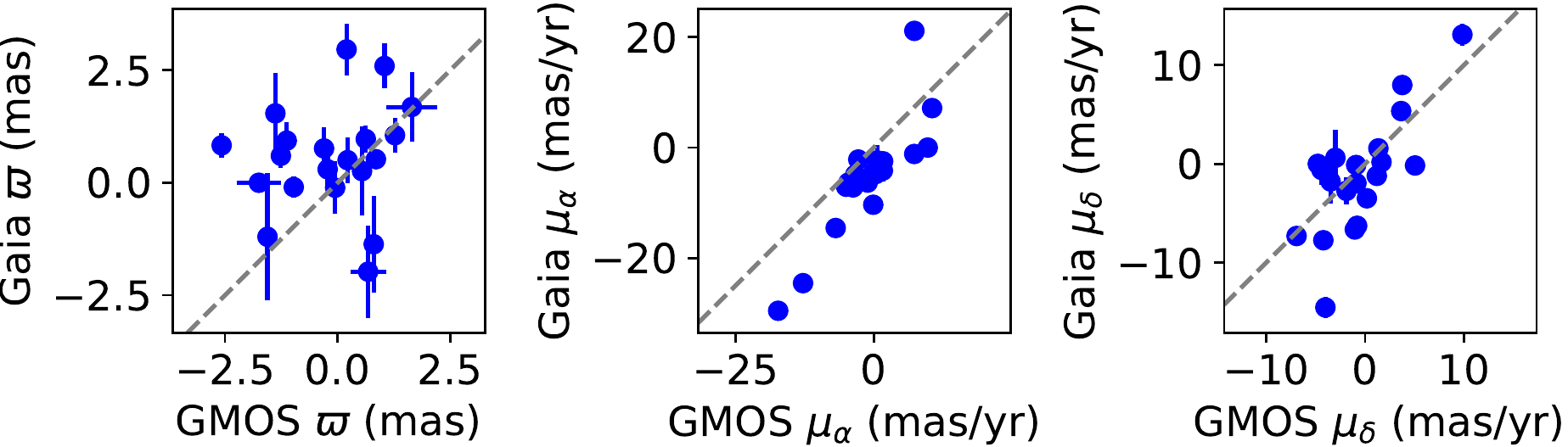}
\caption{Comparison of parallaxes and proper motions between our determinations and GDR2 in the \tmten field. In this case there are 21 reference stars that are both in GDR2 and are used in the final iteration. These are used to derive the corrections to absolute.}
\label{fig:parallax_correction}
\end{figure}

To correct the relative astrometric parameters to absolute quantities, we crossmatched the reference stars with Gaia DR2 sources for which Gaia determined parallaxes. The differences in parallax and proper motions are combined with weights corresponding to their respective uncertainties to obtain the corrections from our relative proper motions and parallaxes to the GDR2 system.  
Figure \ref{fig:parallax_correction} shows these comparisons in the {\tmten} field and Table \ref{tab:corrections} shows the derived corrections (see also the discussion in Section \ref{sec:gdr2}).

\begin{table}
\caption{Corrections to absolute parallax and proper motion. These offsets need to be added to the relative parameters. The number of used reference satrs is $N_\mathrm{ref}$.}
\label{tab:corrections}
\centering
\begin{tabular}{rrr@{$\pm$}lr@{$\pm$}lr@{$\pm$}l}
\hline
\hline
Object & $N_\mathrm{ref}$ & \multicolumn{2}{c}{$\Delta\varpi$} & \multicolumn{2}{c}{$\Delta\mu_{\alpha^\star}$} & \multicolumn{2}{c}{$\Delta\mu_\delta$} \\
& & \multicolumn{2}{c}{(mas)} & \multicolumn{2}{c}{(mas/yr)} & \multicolumn{2}{c}{(mas/yr)} \\

\hline
\tmten & 21 & $0.82$ & $0.11$ & $-5.64$ & $0.52$ & $-0.12$ & $0.39$ \\
\tmeight & 9 & $0.53$ & $0.11$ & $1.48$ & $0.77$ & $-0.75$ & $0.35$ \\
\hline
\end{tabular}
\end{table}

\subsection{Fitting the standard model of parallaxes and proper motions}
{In all of the following analyses we are using all the individual frame data for the model fitting. For better visualisation of the results, however, we display only the epoch averages in the figures.}

\begin{figure}
\centering
\includegraphics[width=0.8\linewidth]{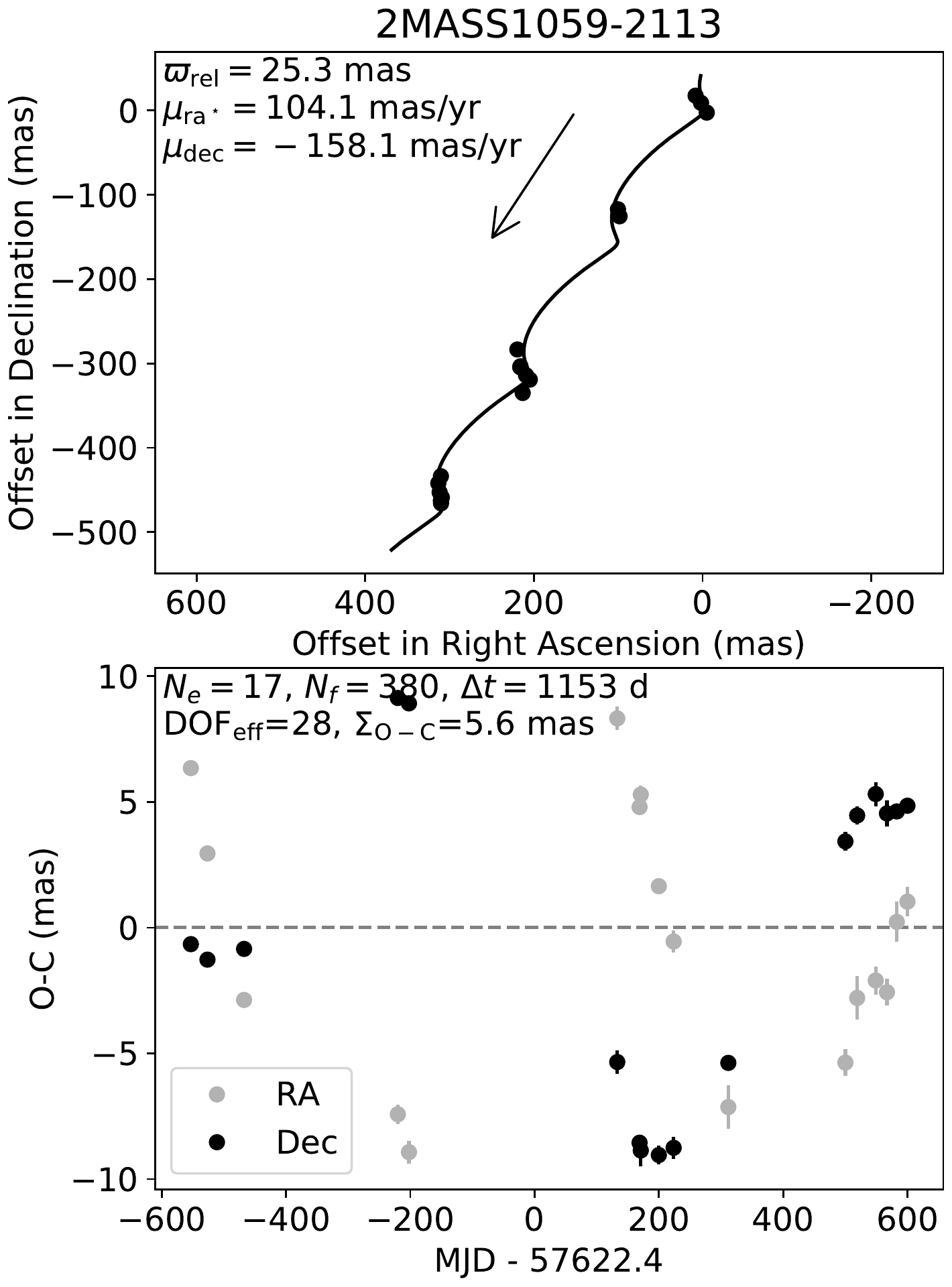}
\caption{Results of fitting the 6-parameter model (PPM+DCR) to \tmten. The on-sky motion (top) and the epoch residuals (bottom) are shown. Significant excess signal is evident.}
\label{fig:ppm_2M1059}
\end{figure}

\begin{figure}
\centering
\includegraphics[width=0.8\linewidth]{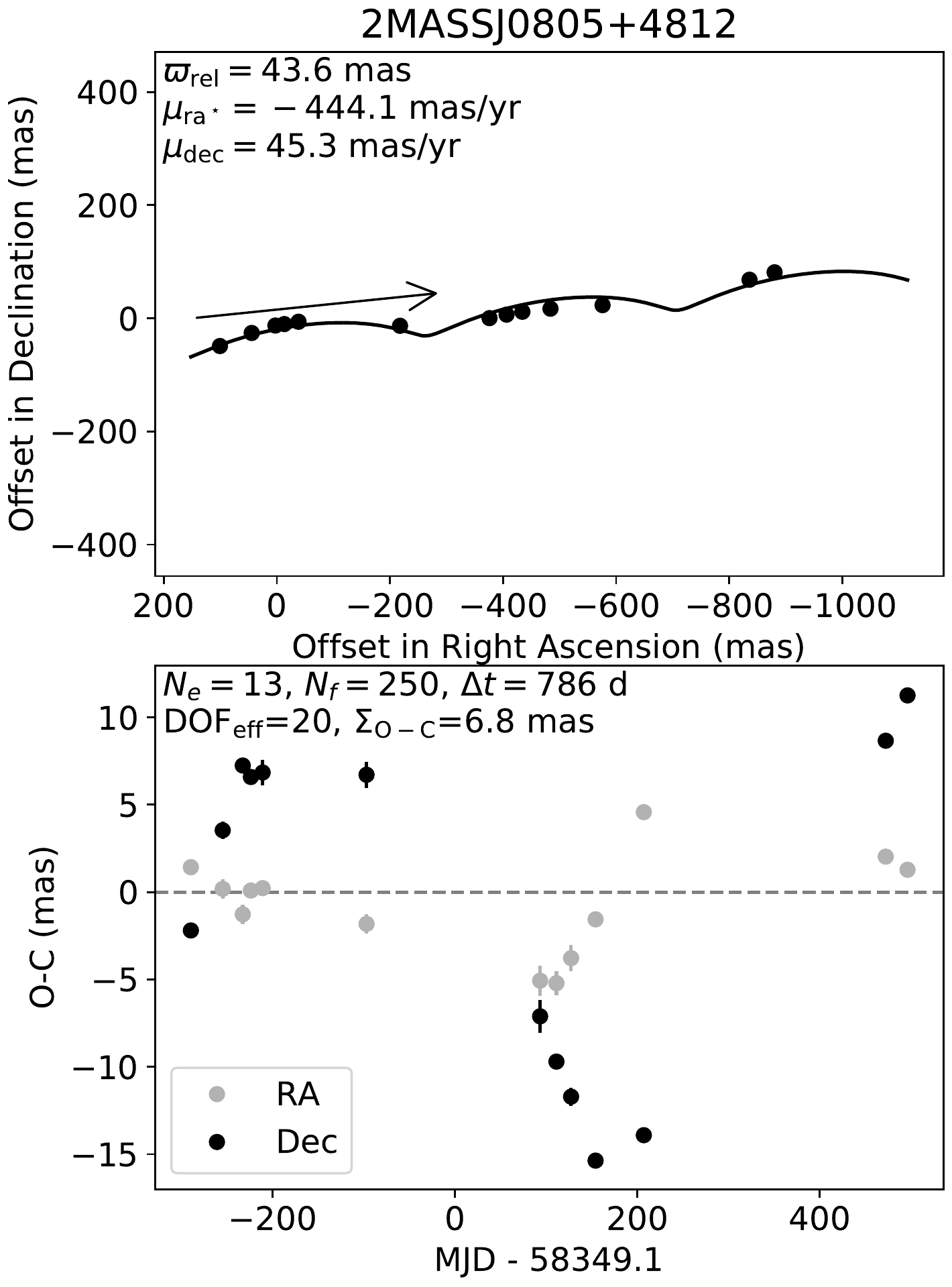}
\caption{Results of fitting the 6-parameter model (PPM+DCR) to \tmeight.}
\label{fig:ppm_2M0805}
\end{figure}

Upon convergence of the iterative astrometric fitting procedure, we adjusted the standard linear model Eq.\ (\ref{eq:axmodel}) to the relative positions of the target in the reference field.
Figures \ref{fig:ppm_2M1059} and \ref{fig:ppm_2M0805} show the results of fitting this 6-parameter model to the astrometry data. Very large excess residual noise is detected for both sources, as expected for the expected binary astrometric motion of our targets.

\subsection{Modelling the target's orbital photocentre motion and radial velocity}\label{sec:modelling}

The Keplerian orbit model adds an additional seven free parameters to the relative offsets $\Delta\alpha^{\star}$ and $\Delta\delta$ of the target's position. These are the eccentricity $e$, the argument of periastron $\omega$, the orbital period $P$, the longitude of ascending node $\Omega$, the orbital inclination $i$, the time of periastron passage $T_\mathrm{P}$, and the semi-major axis of the photocentre orbit $\alpha$.
We also include the astrometric nuisance offset parameters $s_\alpha$ and $s_\delta$ \citep{Sahlmann:2013ab}.

For the spectral binaries in our sample, we estimated the magnitude difference between the components in the filter bandpass, which allows us to relate the photocentre orbit size to the barycentre orbit size $a_1$ of the primary, as described in Section \ref{sec:targets}. 
The relative $i$-band magnitude differences (Table~\ref{tab:targets}) were inferred from the relative 2MASS $J$-band magnitude differences estimated from spectral decomposition analysis \citep{2016ApJ...827...25B,Bardalez-Gagliuffi:2014aa_} and component $i-J$ colors based on the emprical color/spectral type relations of \citet{2016A&A...589A..49S}. Uncertainties in the color relation, component spectral types, and relative $J$ magnitudes (which dominated the error budget) were propagated to compute the relative $i$-band magnitude difference uncertainty.
In addition, we established prior estimates for the primary mass $M_1$ from a population synthesis simulation. We generated 10$^4$ simulated ultracool dwarfs assuming a constant age distribution, a mass distribution $\frac{dN}{dM} \propto M^{-0.5}$ over 0.01 $\leq$ M $\leq$ 0.15, evolutionary models from \citet{Martin:2003ly}, and an empirical temperature/spectral type mapping from \citet{2013ApJS..208....9P}. We sampled the masses of all sources with simulated spectral types within $\pm$0.5 subtypes of the inferred primary component. Prior knowledge on $M_1$ allows us to directly fit the companion mass $M_2$ instead of $\alpha$.

\begin{figure}
\includegraphics[width=\linewidth]{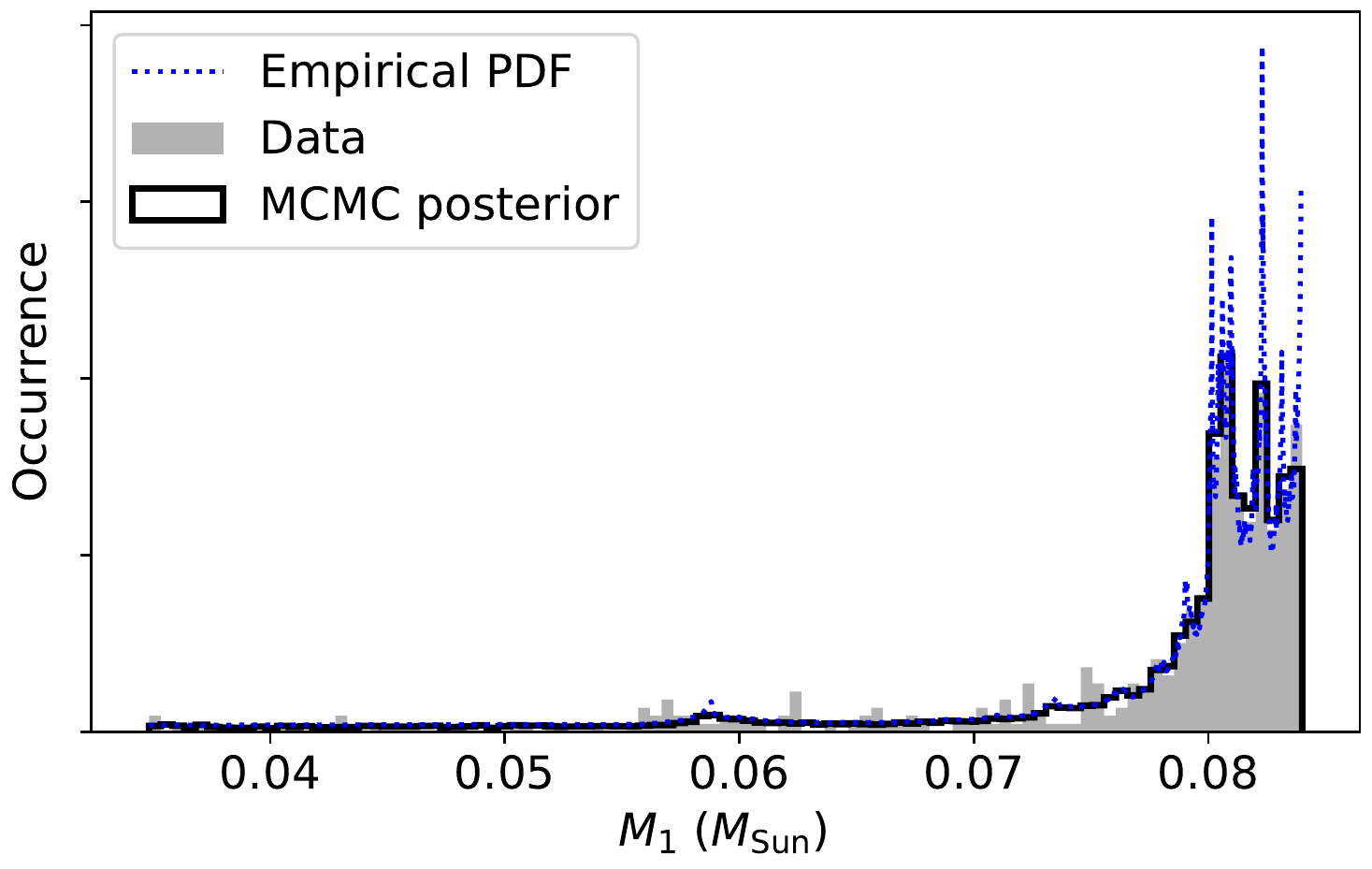}
\caption{Primary mass distribution used for \tmten.}
\label{fig:m1_tmten}
\end{figure}

The resulting primary mass distribution for \tmten\ is shown in Figure \ref{fig:m1_tmten}. We approximated this distribution with the empirical probability distribution function (dotted blue line), which we used to implement a primary mass prior for the MCMC. The resulting posterior distribution is shown with a solid black line and reproduces the input samples well. We implemented the same process for \tmeight.

To mitigate the effects of correlations and limitations that naturally exists for certain orbital parameters, we transform between the following combinations as needed: 
$\lambda_{\rm Ref} \Longleftrightarrow T_\mathrm{P}$, where $\lambda_{\rm Ref}$ is the mean longitude at time $T_\mathrm{Ref}$;  ($M_2$, $i$) $\Longleftrightarrow$ ($\sqrt{M_2} \sin{i}$, $\sqrt{M_2} \cos{i}$); and ($e$, $\omega$) $\Longleftrightarrow$ ($\sqrt{e}\sin\omega$, $\sqrt{e}\cos\omega$)

The radial velocity of the binary is fully characterised by the parameters above with the addition of the systemic velocity $\gamma$.

\subsection{Search for and characterisation of orbital motion}
Since the standard model for astrometric motion does not fit the data well, we searched for orbital motion for all targets using the methods described in \cite{Sahlmann:2013ab}: A Genetic Algorithm is employed to efficiently sample the allowed ranges of non-linear parameters ($P$, $e$, $T_P$), and as the algorithm evolves the regions of minimum $\chi^2$ are determined. 

An inherent parameter degeneracy exists when determining orbital parameters from astrometric data, in that the two solutions with ($\omega$, $\Omega$) and ($\omega+180\degr$, $\Omega+180\degr$) are indistinguishable. We will call the second solution the `degenerate` orbit.
The two can be disentangled when RV data are available because the RV signature of the degenerate orbit is inverted about the systemic velocity, i.e.\ RV$_\mathrm{degenerate}= -\mathrm{RV}_\mathrm{nominal} + \gamma$.
 
The best-fit parameters determined by the Genetic Algorithm were used as starting values for a Markov Chain Monte Carlo (MCMC) analysis similar to that described in \citet{Sahlmann:2016ab}.
We used the \texttt{emcee} package \citep{Foreman-Mackey:2013aa} to implement the MCMC and expressed the binary model using  \texttt{pystrometry} \citep{johannes_sahlmann_2019_3515526} \footnote{\url{https://github.com/Johannes-Sahlmann/pystrometry}} with the parameter vector $\teta$ composed of $\Delta \alpha^\star_0$, $\Delta \delta_0 $, $\varpi$, $\mu_{\alpha^\star}$, $\mu_\delta$, $\rho$,  $P$, $\sqrt{e}\sin\omega$, $\sqrt{e}\cos\omega$, $\lambda_{\rm Ref}$,  $s_\alpha$, $s_\delta$, $M_\mathrm{2} \sin{i}$, $M_\mathrm{2} \cos{i}$, and $\Omega$. We added the systemic velocity $\gamma$ when RV data were included.
The magnitude difference $\Delta \mathrm{mag}$ and the parallax correction $\Delta \varpi$ are incorporated as Gaussian priors in the MCMC \citep[see][]{Sahlmann:2016ab}, whereas the primary mass $M_1$ prior was implemented as described in the previous section.
Finally, the reference time $T_\mathrm{Ref}$ and the absolute coordinates enter the model as constants. 

\section{Results}

\subsection{The orbit of \tmten}
\begin{figure}
\includegraphics[width=\linewidth, trim=0mm 0mm 0mm 6.3mm, clip]{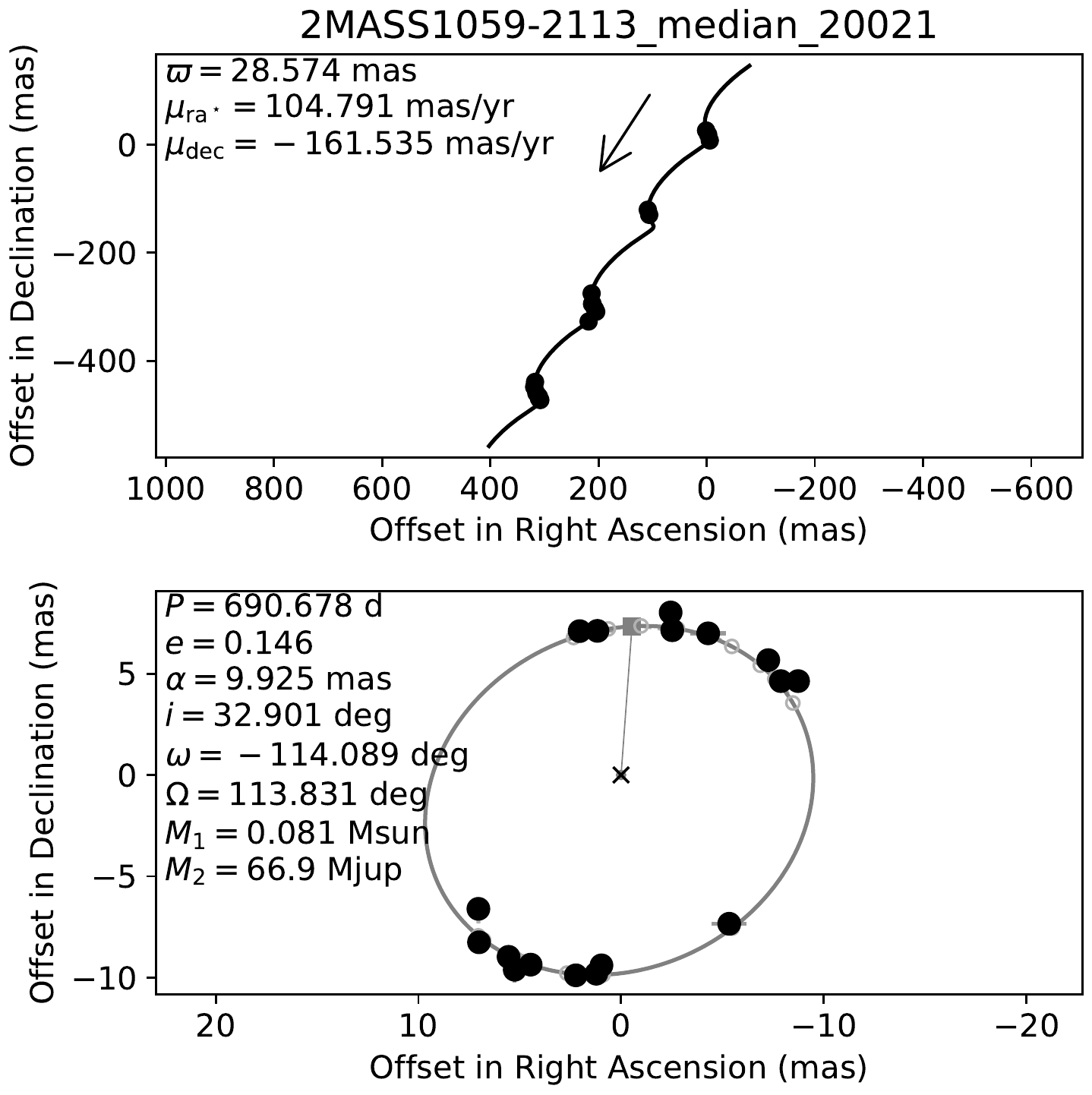}
\caption{\emph{Top}: Sky-projected motion of \tmten\ measured with GMOS-S after removal of orbital motion and chromatic refraction. \emph{Bottom}: Sky-projected photocentre orbit. The cross identifies the barycentre and the gray square shows the periastron location. Epoch{-averaged} measurements are shown with black circles and the `median model` is shown by the curve. {The individual frame measurement that are actually used in the analysis are not shown for clarity.}}
\label{fig:orbit_tmten}
\end{figure}

\begin{figure}
\includegraphics[width=\linewidth, trim=0mm 0mm 0mm 0mm, clip]{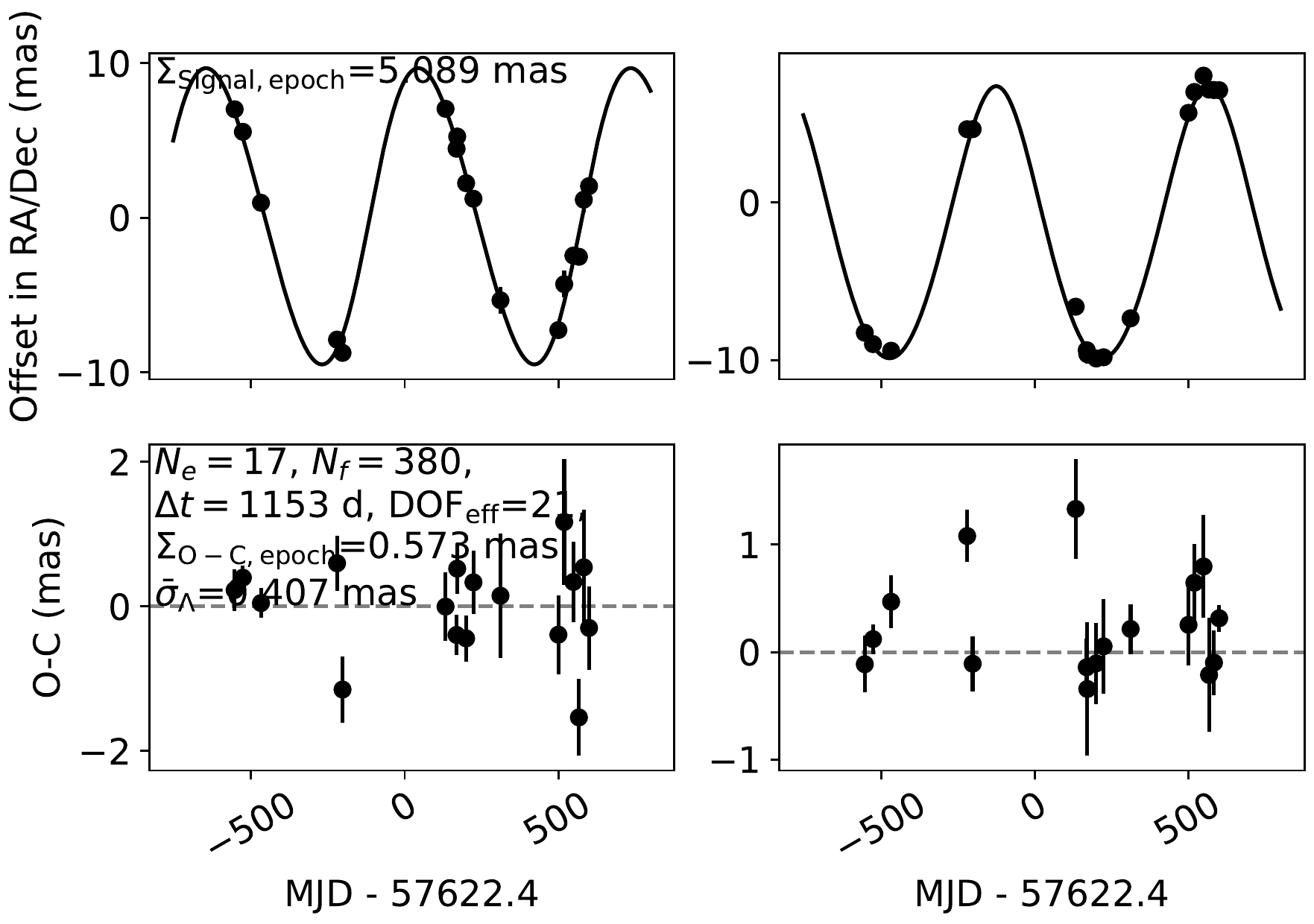}
\caption{The top panels show the orbital motion of \tmten\ in RA (left) and Dec (right) as a function of time with the epoch{-averaged} residuals at the bottom. {Individual-frame measurements are not shown for clarity.}}
\label{fig:orbit_time_tmten}
\end{figure}

\begin{figure}
\includegraphics[width=\linewidth]{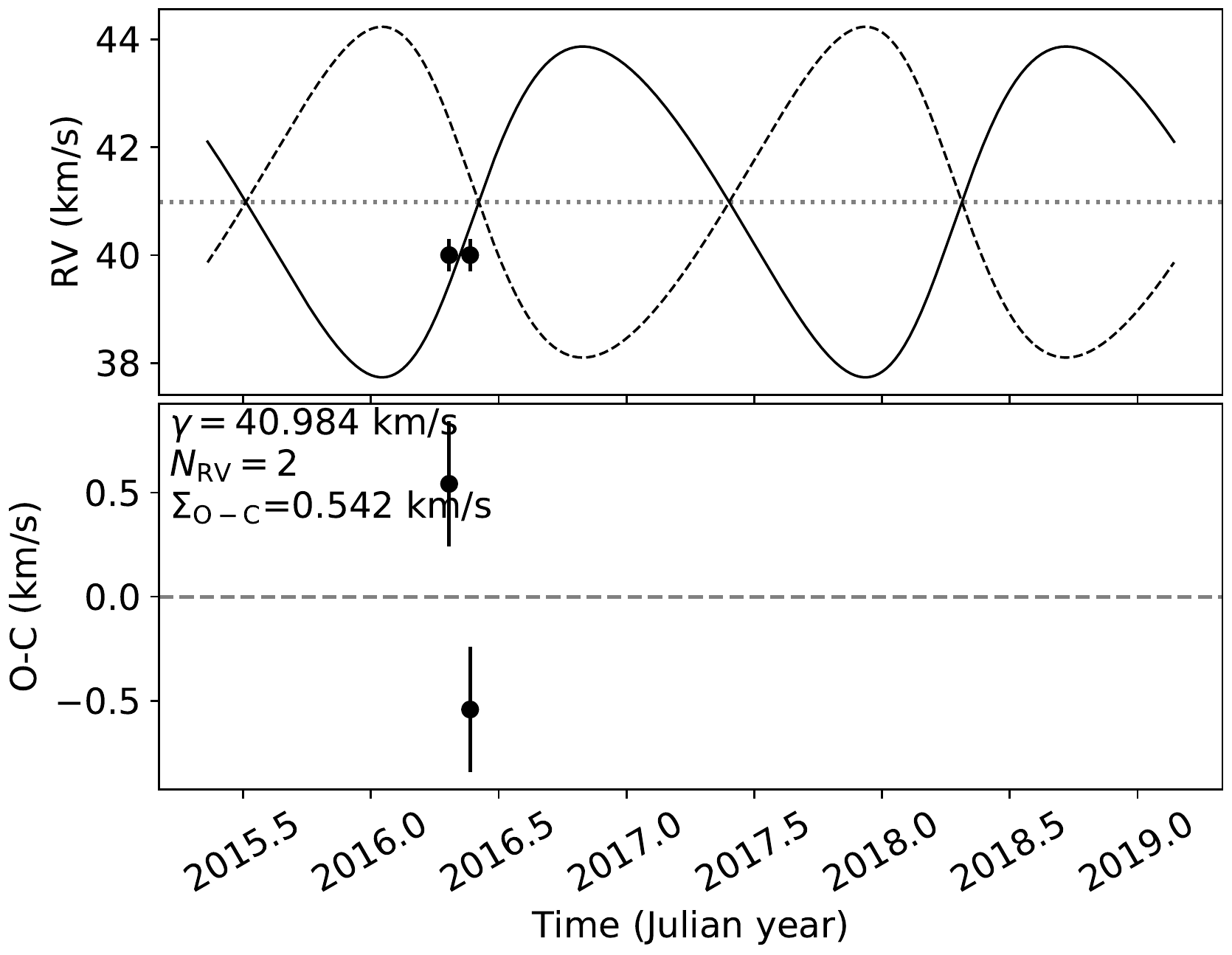}
\caption{\emph{Top}: The predicted radial velocity curve of \tmten\ according to our solution (solid line) and the degenerate solution with $\omega+180\degr$ and $\Omega+180\degr$ (dashed line). \emph{Bottom}: The residual after fitting for the systemic velocity as single parameter while keeping all other parameters fixed.}
\label{fig:orbit_rv_tmten}
\end{figure}

The Genetic Algorithm identified a unique solution that corresponds to a low-eccentricity orbit with a period of $\sim$690 days. The subsequent MCMC analysis provided well-constrained parameters with fast chain convergence and small parameter correlations. 
Table \ref{tab:param_tmten} lists the adopted solution parameters determined as the median of the posterior distributions with 1-$\sigma$-equivalent confidence intervals. Figures \ref{fig:orbit_tmten} and \ref{fig:orbit_time_tmten} show the measured astrometric motion, the fitted Keplerian orbit, and the residuals of the model. The predicted RV orbit is shown in Figure \ref{fig:orbit_rv_tmten} together with the two available RV measurements (Table \ref{tab:frame}), which allowed us to determine a systemic velocity of $\gamma = 40.98 \pm 0.54$ km/s. We cannot use these RVs to distinguish between the nominal and degenerate solution, and for the degenerate orbit with $\omega+180\degr$ and $\Omega+180\degr$ we determine an alternative systemic velocity of  $\gamma' = 39.06 \pm 0.54$ km/s. 

Our results unambiguously confirm the binary nature of {\tmten} and determine its orbital solution for the first time. 
The measured absolute parallax places the system at a distance of 35 pc. The orbit is moderately eccentric ($e\simeq0.14$) and the 1.9 year period corresponds to a relative semimajor axis of 0.8 AU. For a primary mass of $0.082^{+0.002}_{-0.009}M_{\odot}$ we determine the mass of the T3-dwarf companion at $0.064^{+0.004}_{-0.005} M_{\odot}$.

\begin{table}
    \caption{MCMC posterior parameters of \tmten. The parameter $M_\mathrm{tot}$ indicates the total system mass.}
    \label{tab:param_tmten}
    \centering
    \begin{tabular}{lr}
    \hline
    \hline
    Parameter & Value \\
    \hline
    
$\Delta \alpha^\star_0$ (mas) & $152.30^{+1.46}_{-1.37}$ \\[3pt]
$\Delta \delta_0 $ (mas) & $-203.80^{+0.91}_{-0.86}$ \\[3pt]
$\varpi_\mathrm{abs}$ (mas) & $28.57^{+0.56}_{-0.61}$ \\[3pt]
$\mu_{\alpha^\star}$ (mas yr$^{-1}$) & $104.79^{+0.20}_{-0.20}$ \\[3pt]
$\mu_\delta$ (mas yr$^{-1}$) & $-161.53^{+0.11}_{-0.11}$ \\[3pt]
$\rho$ (mas) & $-26.34^{+0.35}_{-0.33}$ \\[3pt]
$P$ (day) & $690.68^{+3.41}_{-3.59}$ \\[3pt]
$P$ (yr) & $1.891^{+0.009}_{-0.010}$ \\[3pt]

$\Omega$ ($^\circ$) & $113.83^{+7.43}_{-7.64}$ \\[3pt]
$\lambda_\mathrm{ref}$ ($^\circ$) & $-51.84^{+8.31}_{-8.06}$ \\[3pt]
$\sqrt{e}\sin\omega$ () & $-0.29^{+0.09}_{-0.07}$ \\[3pt]
$\sqrt{e}\cos\omega$ () & $-0.15^{+0.32}_{-0.21}$ \\[3pt]
$\sqrt{M_2}\sin{i}$ ($M_\mathrm{Jup}$) & $4.43^{+0.45}_{-0.53}$ \\[3pt]
$\sqrt{M_2}\cos{i}$ ($M_\mathrm{Jup}$) & $6.88^{+0.23}_{-0.30}$ \\[3pt]
$s_\alpha$ (mas) & $1.59^{+0.09}_{-0.08}$ \\[3pt]
$s_\delta$ (mas) & $1.24^{+0.07}_{-0.06}$ \\[3pt]
$e$ () & $0.146^{+0.074}_{-0.053}$ \\[3pt]
$\omega$ ($^\circ$) & $-114.09^{+59.26}_{-29.14}$ \\[3pt]
$i$ ($^\circ$) & $32.90^{+2.94}_{-3.44}$ \\[3pt]
$T_P$ (day) & $57502.49^{+110.81}_{-52.28}$ \\[3pt]

$\alpha$ (mas) & $10.00^{+0.29}_{-0.26}$ \\[3pt]
$a_1$ (mas) & $10.22^{+0.30}_{-0.26}$ \\[3pt]
$a_\mathrm{rel}$ (mas) & $22.90^{+0.44}_{-0.69}$ \\[3pt]
$a_\mathrm{rel}$ (AU) & $0.80^{+0.01}_{-0.02}$ \\[3pt]
$M_2$ ($M_\mathrm{Jup}$) & $66.95^{+4.41}_{-4.84}$ \\[3pt]
$M_2$ ($M_\mathrm{Sun}$) & $0.064^{+0.004}_{-0.005}$ \\[3pt]
$M_\mathrm{tot}$ ($M_\mathrm{Sun}$) & $0.145^{+0.005}_{-0.011}$ \\[3pt]

\multicolumn{2}{c}{Priors}\\
$M_1$ ($M_\mathrm{Sun}$) & $0.081^{+0.002}_{-0.008}$ \\[3pt]
$\Delta\varpi$ (mas) & $0.82^{+0.11}_{-0.11}$ \\[3pt]
$\Delta$mag & $5.08^{+0.33}_{-0.32}$ \\[3pt]
\hline
    \end{tabular}
\end{table}

\clearpage
\subsection{The orbit of \tmeight}
The RV orbit of {\tmeight} was previously determined by \cite{Burgasser:2016aa} to have an orbital period 2.02$\pm$0.03 years and eccentricity of 0.46$\pm$0.05.
Even when left unconstrained by radial motion, the Genetic Algorithm applied to our astrometry of {\tmeight}  identified a unique solution with a well-matched orbital period and eccentricity.

The MCMC for {\tmeight} was implemented with a simultaneous fit to both astrometry and RV data, where we used the three additional RV measurements listed in Table~\ref{tab:rv} along with values from  \cite{Burgasser:2016aa}.

\begin{figure}
\includegraphics[width=\linewidth, trim=0mm 0mm 0mm 0mm, clip]{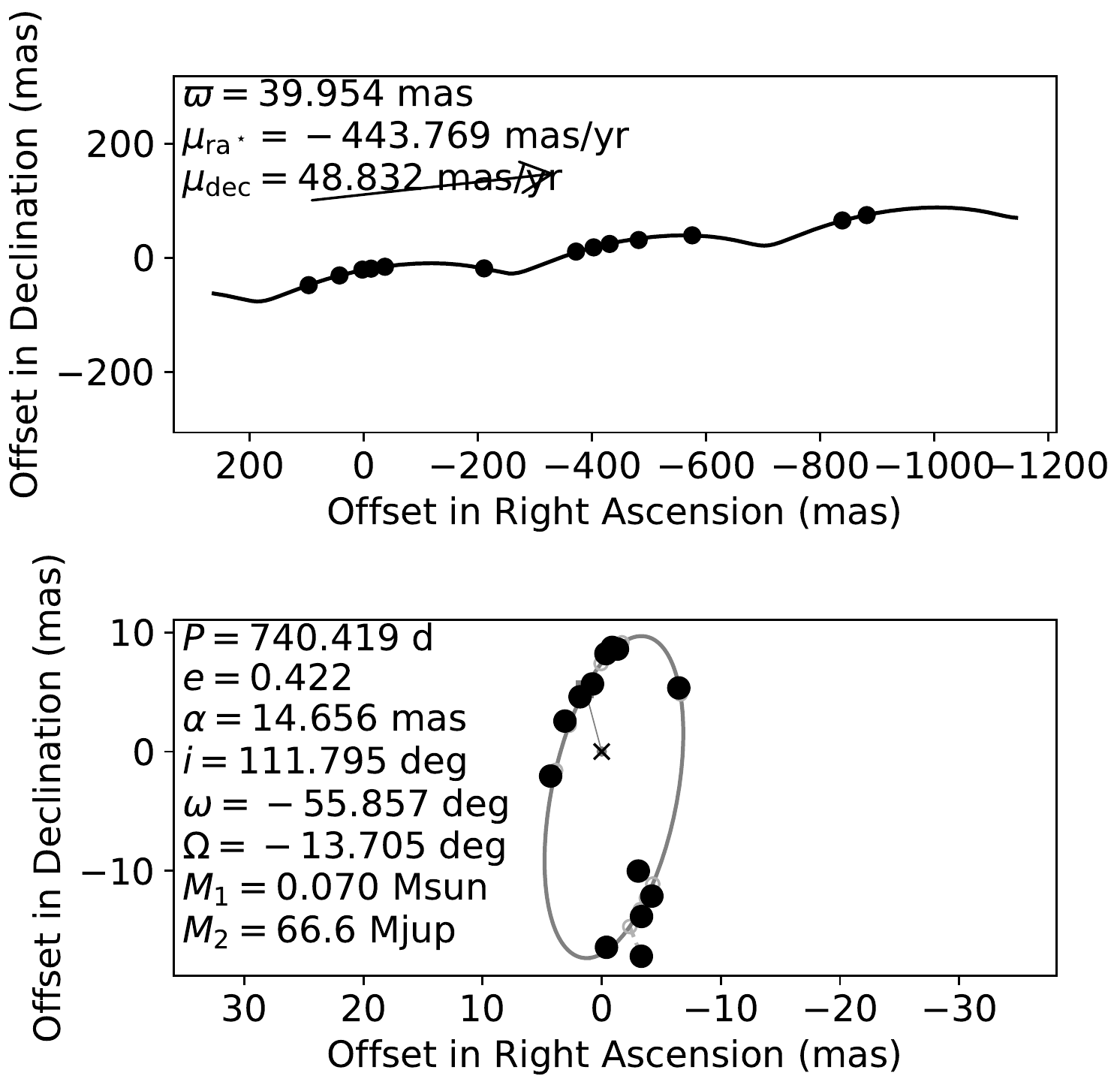}
\caption{Sky-projected motion of \tmeight\ measured with GMOS-N, cf.\ Figure \ref{fig:orbit_tmten}.}
\label{fig:orbit_tmeight}
\end{figure}

\begin{figure}
\includegraphics[width=\linewidth, trim=0mm 0mm 0mm 7.3mm, clip]{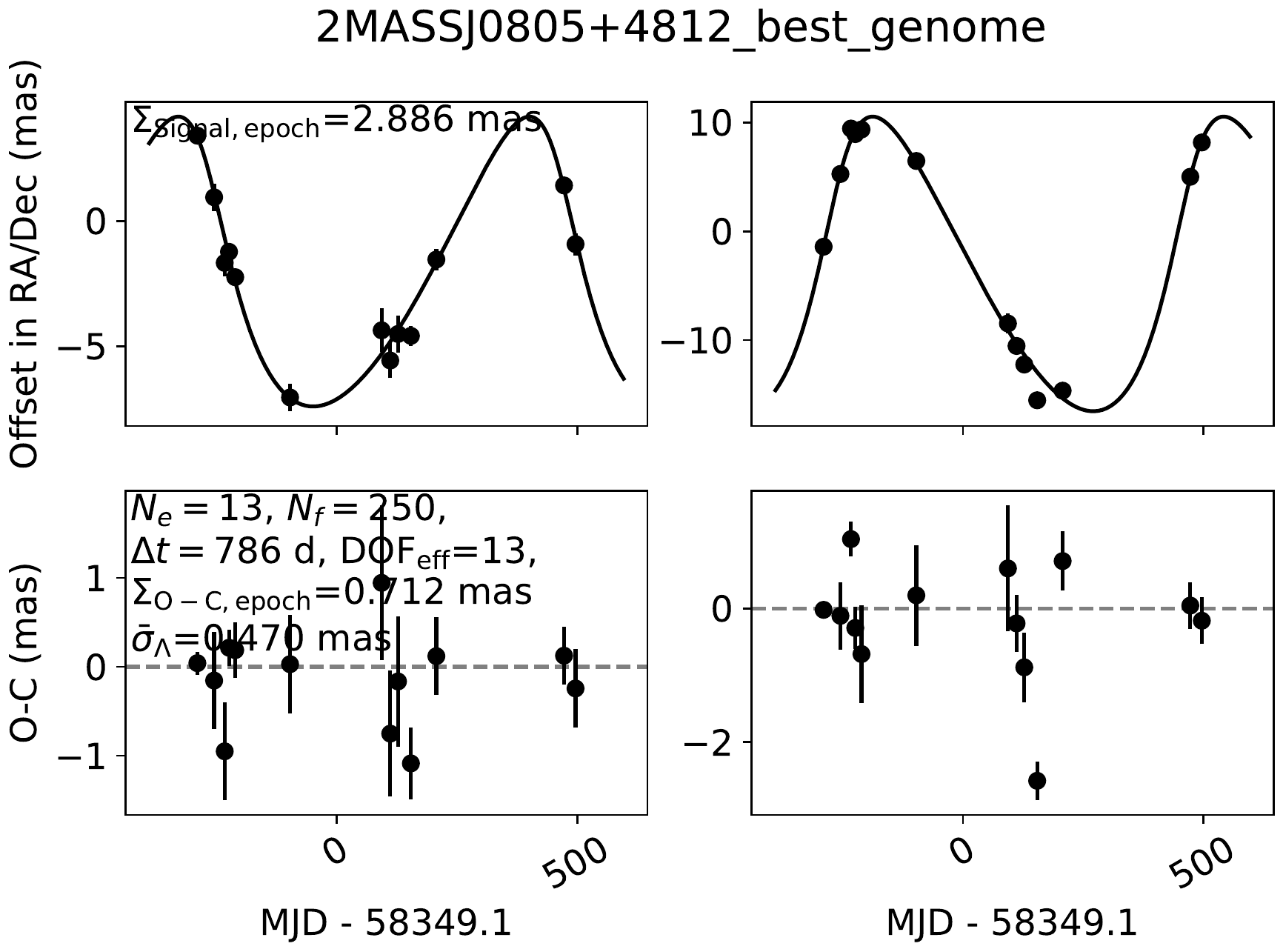}
\caption{Orbital motion of \tmeight as a function of time, cf.\ Figure \ref{fig:orbit_time_tmten}}
\label{fig:orbit_time_tmeight}
\end{figure}

\begin{figure}
\includegraphics[width=\linewidth]{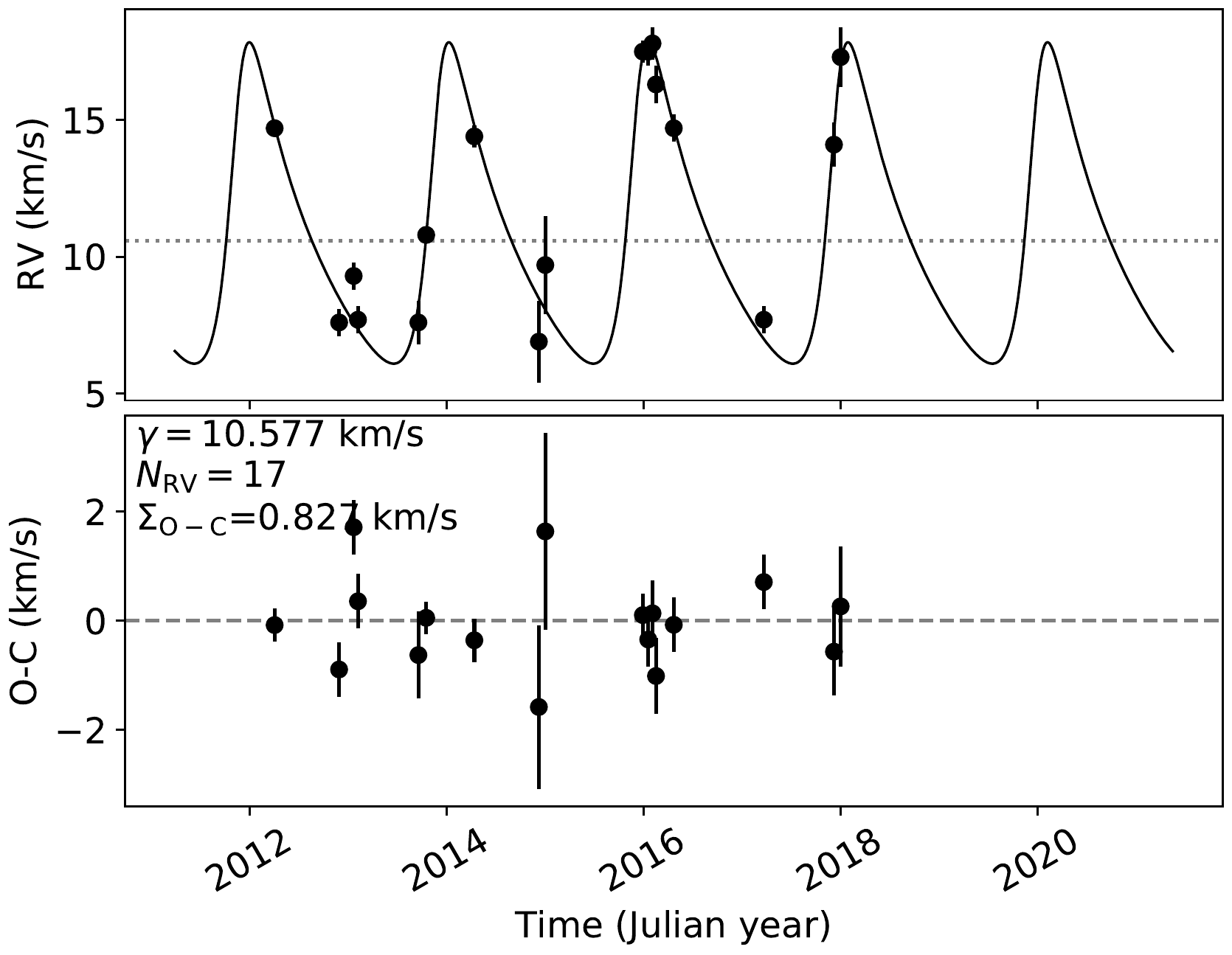}
\caption{RV orbit of \tmeight. The curve in the top panel shows the model resulting from the simultaneous fit to astrometry and RV data. The bottom panel shows the residuals.}
\label{fig:orbit_rv_tmeight}
\end{figure}

Again, this analysis revealed well-constrained parameters with fast chain convergence and small parameter correlations. 
Table \ref{tab:param_tmeight} lists the adopted solution parameters.  Figures \ref{fig:orbit_tmeight} and \ref{fig:orbit_time_tmeight} show the measured astrometric motion, the fitted Keplerian orbit, and the residuals of the model. The RV orbit is shown in Figure \ref{fig:orbit_rv_tmeight}. The direct use of RV data allowed us to break the degeneracy in $\omega$ and $\Omega$, hence this is the unique solution. 

We confirm the period and eccentricity determinations of \cite{Burgasser:2016aa} and the incorporation of astrometric and addition RV data leads to much tighter constraints on these parameters. By determining the astrometric orbit of \tmeight\ for the first time, we also confirm their prediction of a nearly edge-on orbit with an inclination of $112\pm2 \degr$.

\begin{table}
    \caption{MCMC posterior parameters of \tmeight.}
    \label{tab:param_tmeight}
    \centering
    \begin{tabular}{lr}
    \hline
    \hline
    Parameter & Value \\
    \hline
$\Delta \alpha^\star_0$ (mas) & $-293.19^{+0.42}_{-0.41}$ \\[3pt]
$\Delta \delta_0 $ (mas) & $-5.05^{+0.78}_{-0.78}$ \\[3pt]
$\varpi_\mathrm{abs}$ (mas) & $39.91^{+0.35}_{-0.34}$ \\[3pt]
$\mu_{\alpha^\star}$ (mas yr$^{-1}$) & $-443.76^{+0.29}_{-0.28}$ \\[3pt]
$\mu_\delta$ (mas yr$^{-1}$) & $48.82^{+0.31}_{-0.32}$ \\[3pt]
$\rho$ (mas) & $35.06^{+0.61}_{-0.64}$ \\[3pt]
$\gamma$ (m s$^{-1}$) & $10574.07^{+135.69}_{-140.20}$ \\[3pt]
 
$P$ (day) & $740.43^{+1.57}_{-1.63}$ \\[3pt]
$P$ (yr) & $2.027^{+0.004}_{-0.004}$ \\[3pt]

$\Omega$ ($^\circ$) & $-13.69^{+2.02}_{-2.00}$ \\[3pt]
$\lambda_\mathrm{ref}$ ($^\circ$) & $-294.65^{+1.82}_{-1.71}$ \\[3pt]
$\sqrt{e}\sin\omega$ () & $-0.54^{+0.04}_{-0.04}$ \\[3pt]
$\sqrt{e}\cos\omega$ () & $0.37^{+0.05}_{-0.05}$ \\[3pt]
$\sqrt{M_2}\sin{i}$ ($M_\mathrm{Jup}$) & $7.54^{+0.32}_{-0.83}$ \\[3pt]
$\sqrt{M_2}\cos{i}$ ($M_\mathrm{Jup}$) & $-2.98^{+0.36}_{-0.27}$ \\[3pt]
$s_\alpha$ (mas) & $1.82^{+0.13}_{-0.11}$ \\[3pt]
$s_\delta$ (mas) & $1.92^{+0.13}_{-0.12}$ \\[3pt]
$e$  & $0.423^{+0.019}_{-0.019}$ \\[3pt]
$\omega$ ($^\circ$) & $-55.79^{+5.36}_{-5.36}$ \\[3pt]
$i$ ($^\circ$) & $111.85^{+1.55}_{-1.52}$ \\[3pt]
$T_P$ (day) & $58840.28^{+9.17}_{-9.19}$ \\[3pt]

$\alpha$ (mas) & $14.76^{+0.38}_{-0.38}$ \\[3pt]
$a_1$ (mas) & $15.67^{+0.39}_{-0.39}$ \\[3pt]
$a_\mathrm{rel}$ (mas) & $32.58^{+1.01}_{-3.69}$ \\[3pt]
$a_\mathrm{rel}$ (AU) & $0.82^{+0.02}_{-0.09}$ \\[3pt]
$M_2$ ($M_\mathrm{Jup}$) & $66.28^{+5.18}_{-14.04}$ \\[3pt]
$M_2$ ($M_\mathrm{Sun}$) & $0.063^{+0.005}_{-0.013}$ \\[3pt]
$M_\mathrm{tot}$ ($M_\mathrm{Sun}$) & $0.134^{+0.011}_{-0.038}$ \\[3pt]   

\multicolumn{2}{c}{Priors}\\
$\Delta$mag  & $3.81^{+0.07}_{-0.07}$ \\[3pt]
$M_1$ ($M_\mathrm{Sun}$) & $0.069^{+0.008}_{-0.027}$ \\[3pt]
$\Delta\varpi$ (mas) & $0.54^{+0.11}_{-0.11}$ \\[3pt]

\hline
    \end{tabular}
\end{table}

\section{Discussion}
\subsection{Masses of T dwarf companions}
Our astrometric follow-up allowed us to set tight constraints on the masses of the T dwarf companions of \tmten\ and \tmeight. Table \ref{tab:masses} summarises those mass determinations. 
In the context of other dynamical T dwarf masses (e.g. Table 6 of \citet{Dupuy:2019aa}), we see that the mass of \tmeight B is almost equal to the one of \wiseven B estimated by \citet{Dupuy:2019aa}, which also has the same spectral type{, yet a mass higher than the other three T5 dwarfs with measured dynamical masses. From this comparison, \tmeight\ is also a `massive` T dwarf.}

Our mass of $67^{+4}_{-5}\, M_\mathrm{Jup}$  for the T3.5 \tmten B, is {also} significantly higher than its two spectral type equivalents DENIS J2252$-$1730B (T3.5, $41\pm4\, M_\mathrm{Jup}$) and 2MASS J1534$-$2952A (T4.5, $51\pm5\, M_\mathrm{Jup}$).

\begin{figure*}
\includegraphics[width=0.8\textwidth, trim=5mm 0mm 15mm 10mm, clip]{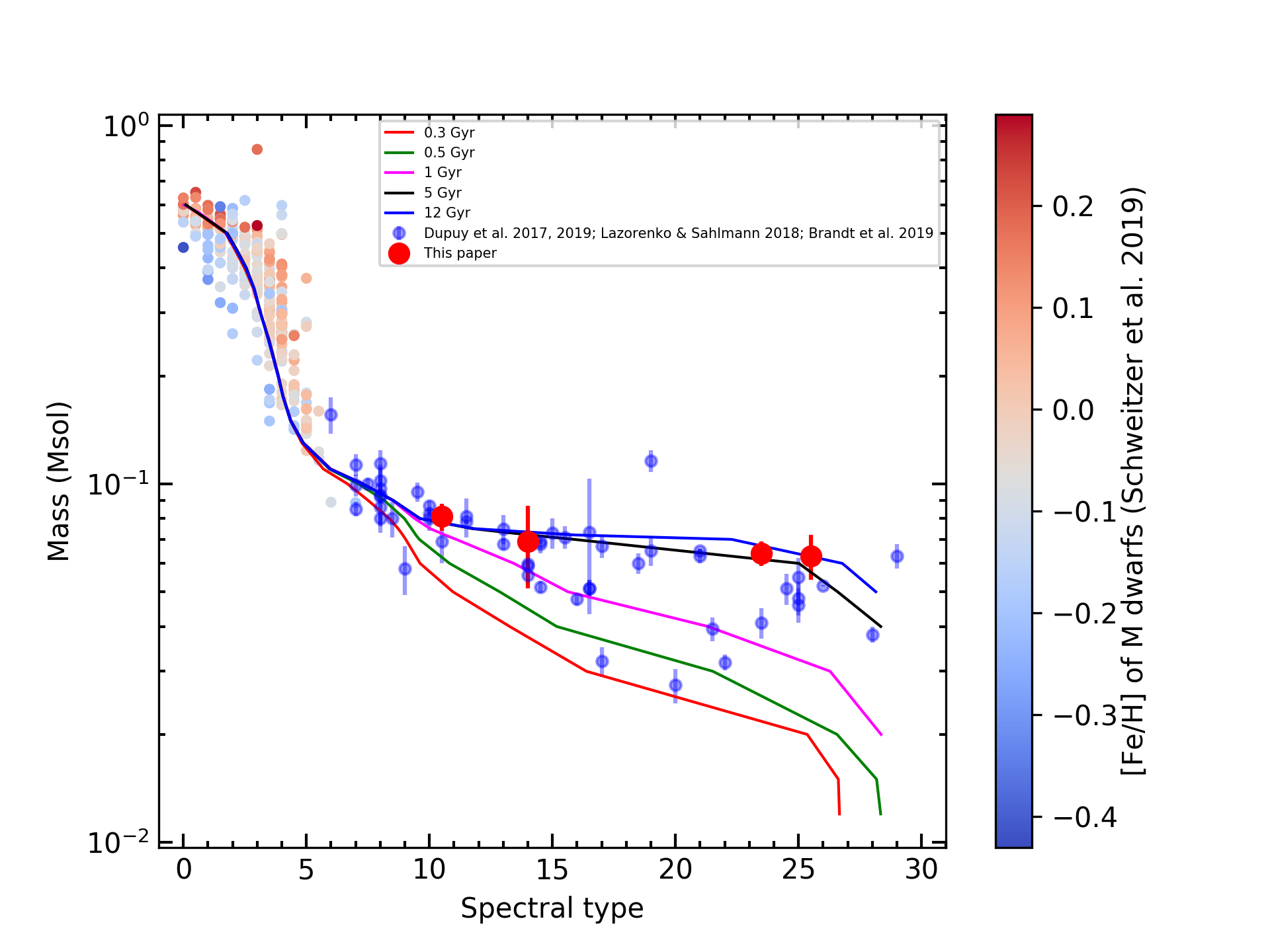}
\caption{Dynamically derived masses in the low luminosity regime as a function of spectral type. On the x-axis, spectral types of M0, L0, and T0 are encoded as 0, 10, and 20, respectively. The \tmeight\ and \tmten\ systems are shown with large red symbols.}
\label{fig:masses}
\end{figure*}

In Figure \ref{fig:masses} we show our results in the context of other low-mass systems with dynamically determined masses, where we included early-to-mid M dwarfs from \citet{2019A&A...625A..68S} and late-M, L, and T dwarfs from \citet{2017ApJS..231...15D, Lazorenko:2018aa, Dupuy:2019aa}. { The curves correspond to isochrones of different ages (0.3, 0.5, 1, 5, and 12 Gyr; \citealt{Baraffe:2015aa}), where we used the spectral type -- effective temperature calibrations of \citet{2019AJ....158...56H}} for M dwarfs and of \citet{Stephens:2009aa} for L and T dwarfs to convert theoretically predicted effective temperatures into spectral types. 

This figure shows that the results of this work are generally consistent with the dynamical masses from other groups in the spectral-type range probed and that \tmeight B is among the latest-type dwarfs with well-constrained masses. The masses derived for each member of our two pairs are compatible with the {5 and 12 Gyr isochrones} at the 1-sigma level and 
{the $>$1 Gyr isochrones show} reasonable agreement with the observational data.

\tmten\ and \tmeight\ join a short yet growing sample of `massive` T dwarfs compared with their predictions from evolutionary models, alongside $\epsilon$~Indi B and C~\citep{2018ApJ...865...28D},  \wiseven~\citep{Dupuy:2019aa}, Gl~229~B~\citep{2019arXiv191001652B}, and HD 4113C~\citep{2018A&A...614A..16C}. Several avenues have been explored to interpret the {high mass of Gl~229~B} within the framework of current evolutionary models, {including unresolved binarity, incorrect astrometry of the primary, low metallicity and atypical old ages ($\gtrsim7\,$Gyr), however none of these hypotheses have been determined as the cause. In particular, activity and kinematic constraints for the age of the primary lead to a broad range of 2-8\,Gyr, but with a higher likelihood at younger ages.} 

 {The T dwarf secondaries of \tmten\ and \tmeight\ also have dynamical masses higher than comparable objects at a given spectral type. These `massive` T dwarfs are marginally consistent with older ages ($\gtrsim7$\,Gyr), yet do not show signatures of low metallicity.  Recent theoretical and computational developments on the equation of state (EOS) describing the interiors of substellar objects, based on improved quantum molecular dynamics calculations in the high density-temperature regime of pressure dissociation and ionization for H and He, can provide a partial solution to this discrepancy. The evolutionary models associated with this EOS lead to more degenerate interiors,  slightly faster cooling rates, cooler temperatures and lower brightness for a given mass \citep{2020arXiv200313717P}.}

\begin{table}
\caption{Dynamical masses of the two T-dwarf companions.}
    \label{tab:masses}
    \centering
\begin{tabular}{lrrrrr}
    \hline
    \hline
Object & Spec. Type & Mass \\
    \hline
\tmten A &     L0.5 &$0.081^{+0.002}_{-0.008} M_\mathrm{Sun}$  \\[3pt]
\tmten B &     T3.5 &$0.064^{+0.004}_{-0.005} M_\mathrm{Sun}$  \\[3pt]
\tmten B &     T3.5 &$67^{+4}_{-5} M_\mathrm{Jup}$  \\[3pt]
\hline
\tmeight A &     L4 & $0.069^{+0.008}_{-0.027} M_\mathrm{Sun}$  \\[3pt]
\tmeight B &     T5.5 &$0.063^{+0.005}_{-0.013} M_\mathrm{Sun}$ \\[3pt]
\tmeight B &     T5.5 &$66^{+5}_{-14} M_\mathrm{Jup}$  \\

    \hline
\end{tabular}
\end{table}

\clearpage

\subsection{Astrometric accuracy achieved with GMOS imaging}
Using GMOS-S and GMOS-N imaging, we achieved epoch residual RMS values of 0.6 mas and 0.7 mas for \tmten\ and \tmeight, respectively. We expect that improved source extraction procedures, better selection of the reference star sample, and more careful treatment of outliers during the iterative astrometry fitting will result in significantly better performance. However, we do not expect to reach the 0.1 mas accuracy performance of our FORS2/VLT program \citep{Sahlmann:2014aa} because the GMOS fields have significantly fewer available reference stars.

\subsection{Comparison with Gaia DR2 parameters}\label{sec:gdr2}
In Table \ref{tab:gdr2_comp} we compare our parallax and proper motion determinations with the values given in the GDR2 catalogue. In all cases, the differences are significant which can primarily be attributed to the GDR2 model that did not account for orbital motion but applied the standard 5-parameter linear model instead. This is also reflected in the elevated GDR2 astrometric excess noise of 2.7 mas and 4.3 mas for \tmten\ and \tmeight, respectively. As a result the GDR2 parameters are biased and we expect out determinations to be more accurate. In particular, our parallax measurements imply significantly larger distances. 

\begin{table}
\caption{Absolute parallaxes and proper motions.}
    \label{tab:gdr2_comp}
    \centering
\begin{tabular}{lrr}
    \hline
    \hline
Parameter & GMOS & GDR2 \\
    \hline
 & \multicolumn{2}{c}{\tmten}\\
$\varpi_\mathrm{abs}$ (mas) & $28.6\pm{0.6}$ & $31.4\pm0.8$ \\
$\mu_{\alpha^\star}$ (mas yr$^{-1}$) & $99.1\pm{0.6}$ & $85.0\pm1.2$ \\
$\mu_\delta$ (mas yr$^{-1}$) & $-161.7\pm{0.4}$ & $-164.2\pm1.0$ \\
  \hline
 & \multicolumn{2}{c}{\tmeight}\\
    $\varpi_\mathrm{abs}$ (mas) & $40.0\pm{0.4}$ & $46.8\pm1.0$ \\
$\mu_{\alpha^\star,\mathrm{abs}}$ (mas yr$^{-1}$) & $-442.3\pm{0.8}$ & $-459.1\pm1.4$ \\
$\mu_{\delta,\mathrm{abs}}$ (mas yr$^{-1}$) & $48.1\pm{0.5}$ & $56.7\pm1.1$ \\
    \hline
\end{tabular}
\end{table}

\section{Conclusions}
We reported the first results of an astrometric follow-up campaign to confirm and characterize spectral binary brown dwarfs. Using both Gemini GMOS imagers, we demonstrated sub-milliarcsecond astrometry and determined the astrometric orbits of \tmten\ and \tmeight\ for the first time, thereby tightly constraining the masses of their T-dwarf companions.  
We showed that astrometric observations represent an underutilised avenue for confirming and characterising the orbits of tight low-mass binaries containing stellar and brown dwarf components. Our survey is particularly efficient because it is guided by prior indications for binarity from near-infrared spectroscopy.

Surveys like ours combined with the ultracool-dwarf binary orbits expected from the Gaia mission and astrometric programs that explore the presence of giant planets around ultracool dwarfs, will lead to the population characterisation of compact ultracool systems over a wide range of mass-ratios.

\section*{Acknowledgments}
This research made use of the databases at the Centre de Donn\'ees astronomiques de Strasbourg (\url{http://cds.u-strasbg.fr}); of NASA's Astrophysics Data System Service (\url{http://adsabs.harvard.edu/abstract\_service.html}); of the paper repositories at arXiv; and of Astropy, a community-developed core Python package for Astronomy \citep{Astropy-Collaboration:2013aa}.

Based on observations obtained at the Gemini Observatory (acquired through the Gemini Observatory Archive and processed using the Gemini IRAF package and gemini\_python), which is operated by the Association of Universities for Research in Astronomy, Inc., under a cooperative agreement with the NSF on behalf of the Gemini partnership: the National Science Foundation (United States), the National Research Council (Canada), CONICYT (Chile), Ministerio de Ciencia, Tecnolog\'{i}a e Innovaci\'{o}n Productiva (Argentina), and Minist\'{e}rio da Ci\^{e}ncia, Tecnologia e Inova\c{c}\~{a}o (Brazil). 

This work has made use of data from the European Space Agency (ESA) mission
{\it Gaia} (\url{https://www.cosmos.esa.int/gaia}), processed by the {\it Gaia}
Data Processing and Analysis Consortium (DPAC,
\url{https://www.cosmos.esa.int/web/gaia/dpac/consortium}). Funding for the DPAC
has been provided by national institutions, in particular the institutions
participating in the {\it Gaia} Multilateral Agreement.

ELM acknowledges funding from Ministerio de Economia y Competitividad and the Fondo Europeo de Desarrollo Regional (FEDER) under grant AYA2015-69350-C3-1-P \@.

\bibliographystyle{mnras}
\bibliography{references} 

\bsp	
\label{lastpage}
\end{document}